\documentclass[onecolumn]{emulateapj}
\usepackage{lscape}
\newcommand{\be}{\begin{equation}}
\newcommand{\ee}{\end{equation}}
\newcommand{\bea}{\begin{eqnarray}}
\newcommand{\eea}{\end{eqnarray}}
\shortauthors{Yan \& Lazarian}
\begin{document}
\title{Polarization of absorption lines as a diagnostics of circumstellar,  
interstellar and intergalactic magnetic fields:  Fine structure atoms}
\author{Huirong Yan\altaffilmark{1,2} \& A. Lazarian\altaffilmark{1}}
\altaffiltext{1}{Department of Astronomy, University of Wisconsin, 475 N. Charter
St., Madison, WI 53706, yan, lazarian@astro.wisc.edu}
\altaffiltext{2}{Present address: 
Canadian Institute for Theoretical Astrophysics, 60 St. George, Toronto, ON M5S3H8, yanhr@cita.utoronto.ca}

\begin{abstract}

The relative population of the fine structure sublevels of an atom's ground state is affected by 
radiative transitions induced by an
anisotropic radiation flux. This causes the 
alignment of atomic angular momentum. In terms of observational
consequences for the interstellar and intergalactic medium,
this results in the polarization of the absorption lines. In the paper 
we consider the conditions necessary for this effect and
provide calculations of polarization from 
a few astrophysically important atoms and ions
 with multiple upper and lower levels for an arbitrary orientation
of magnetic fields to the a) source of optical pumping, b) direction
of observation, c)  absorbed source. We also consider
an astrophysically important ``degenerate'' case when the
source of optical pumping coincides with the source of the absorbed
radiation.  We present analytical expressions that relate 
the degree of linear polarization and
the intensity of absorption to the 3D orientation of the 
magnetic field with respect to the pumping source, the source of 
the absorbed radiation, and the direction of observations. We
discuss how all these parameters can be determined via simultaneous
observations of several absorption lines and suggest graphical means
that are helpful in practical data interpretation.
We prove that studies of absorption line polarization 
provide a unique tool to study 3D magnetic field topology in various
astrophysical conditions. 

\end{abstract}
\keywords{ISM:  atomic processes---magnetic fields---polarization}

\section{Introduction}

Magnetic fields play extremely important roles in many astrophysical circumstances, e.g., the interstellar medium, intergalactic medium and quasars, etc.
Unfortunately,  there are only a few techniques for magnetic field studies
that are available (see below). Each technique is sensitive to magnetic fields in 
particular environments. Therefore even the directions of magnetic field obtained  for
the same region of sky with different techniques differ substantially.
The simultaneous use of different techniques provides a possibility of
magnetic field tomography.

 Polarimetry of aligned dust provides  a way of studying 
magnetic field direction in the diffuse interstellar medium, molecular clouds, 
circumstellar and interplanetary medium (see review by Lazarian 2003).  
Substantial progress in the understanding of grain alignment has been achieved in the last decade making the technique more reliable. The technique is, 
however, challenging to apply to low column densities, as the polarization
signal becomes too weak. It has other limitations when dealing with high 
densities (Cho \& Lazarian 2005).

Polarimetry of some molecular lines using the Goldreich-Kylafis (1982) effect has recently been shown to be a good tool for magnetic field  studies in molecular clouds (see Girart, Crutcher \& Rao 1999). However, the magnetic field direction obtained has an uncertainty of 90 degrees, which may be confusing. The 
technique is most promising for dense CO clouds.

Zeeman splitting (see Crutcher 2004) provides a good way to get magnetic
field strength, but the measurements are very time consuming and only
the strongest magnetic fields are detectable this way 
(see Heiles \& Troland 2004).

Synchrotron emission/polarization as well as Faraday rotation provide an
important means to study magnetic fields either in distinct regions with strong magnetic
fields or over wide expanse of the magnetized diffuse media (Haverkorn 2005).

Here we discuss yet another promising technique to study magnetic fields that employs optical and UV polarimetry. This technique can be used for 
interstellar\footnote{Here interstellar is understood in a 
general sense, which, for instance,
includes refection nebulae.}, and
 intergalactic studies as well as for studies of magnetic fields in
QSOs and other astrophysical objects.  As we discuss below,
the technique is based on
the ability of atoms and ions to be aligned by external radiation in their ground state
and be realigned through precession in magnetic  field. 
Henceforth, we shall not
 distinguish atoms and ions and use word ``atoms'' dealing  with both species.
This is justifiable, as the alignment of ions and atoms does not differ.

The physics of atomic alignment is rather straightforward.
It has been known that atoms can be aligned through interactions with the anisotropic flux
of resonance emission (see review Happer 1972 and references therein). Alignment is  understood here in terms of orientation
of the angular momentum vector
$\bf J$, if we use the language of classical mechanics. In quantum
terms  this means a difference in the population of sublevels corresponding to
projections of angular momentum to the quantization axis. Later
we shall
refer to ${\bf J}$ alignment in the ground state of atoms as the 
``atomic alignment''.  Note that we deal with the ground state as it 
is long lived and therefore sensitive to
weak magnetic field.

It is worth mentioning that atomic alignment was
studied in laboratory in relation with early 
day maser 
research (see Hawkins 1955). This effect was noticed and employed in the interstellar case 
by Varshalovich (1968) in the case of hyperfine structure of the ground state. Varshalovich (1971) suggested that atomic alignment enables us to detect the direction of magnetic fields in the interstellar medium, but did not provide a thorough quantitative study.  

A more rigorous study of using atomic alignment to diagnose weak 
magnetic field in diffuse media was conducted in Landolfi \& Landi Degl'Innocenti (1986). However, in that case, an idealized 
two-level atom was considered. In addition, polarization of emission from this atom  was
discussed for a very restricted geometry of observations,
namely, the magnetic field is along the 
line of sight and both of these directions are perpendicular to the incident light. 
This made it rather difficult to use this study as a tool for practical mapping of magnetic fields in various astrophysical environments.

In this paper, we shall consider the alignment of atoms with fine structure of the ground level. Realistic atomic species will be studied, taking into account their multi-level structure. We shall show that atomic alignment can influence the line polarization and intensity of absorption lines. The polarization dependence on the 3D geometry of both radiation and magnetic field will be provided. A study of the alignment of species
within hyperfine structure of the ground state (e.g. Na, HI) and
the polarization of emission lines
 will be provided in our companion paper.

It should be pointed out that the atomic alignment we deal with in this paper differs from the Hanle effect that solar researcher have studied. Hanle effect is depolarization and rotation of the polarization vector of the resonance scattered lines in the presence of a magnetic field, which happens when the magnetic splitting becomes comparable to the decay rate of the excited state of an atom. The research into emission line polarimetry resulted in important
change of the views on solar chromosphere (see Landi Degl'Innocenti 1983a, 1984, 1998, Stenflo \& Keller 1997, Trujillo
Bueno \& Landi Degl'Innocenti 1997, Trujillo
Bueno 1999, Trujillo Bueno et al. 2002, Manso Sainz \& Trujillo Bueno 2003). However, these studies correspond to a setting different from the one we consider here.
In this paper we concentrate on the weak field regime, in which it is 
the atoms at ground level that are repopulated due to magnetic precession, while the Hanle effect is negligible for the upper state. 
This is the case, for instance, of the interstellar medium. The polarization of absorption lines that we study here is thus more
informative. In many cases, we are in optically thin regime, so we do not need to be concerned about radiative transfer.

In what follows we present the
discussion of basic atomic alignment physics and optical pumping in \S2. Detailed treatment of statistical equilibrium of fine structure atoms in a magnetic field is given in \S3. Following this we then calculate, for a number of elements, their alignment  and the induced polarizations of their emission and absorption lines in \S4. To help our reader understand the logic of the calculations we provide a quantitative level summary of the procedures in \S5. The discussion and the summary are provided in, respectively, \S6 and \S7.  The Appendix contains the information necessary for
understanding our formalism as well as auxiliary calculations.

\section{Conditions for atomic alignment}

\begin{figure}
\includegraphics[%
  width=0.45\textwidth,
  height=0.4\textheight]{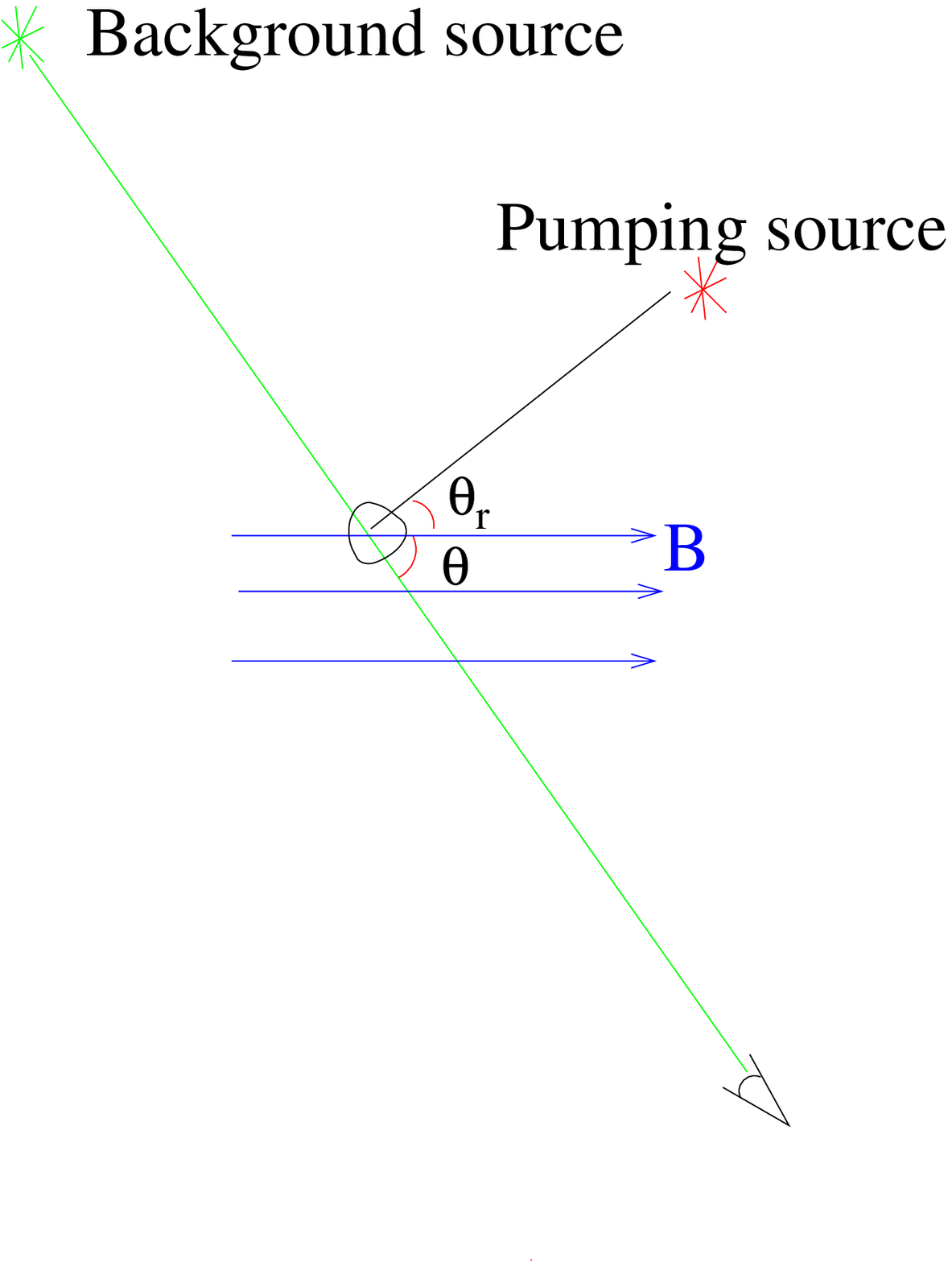}
\includegraphics[%
  width=0.45\textwidth,
  height=0.4\textheight]{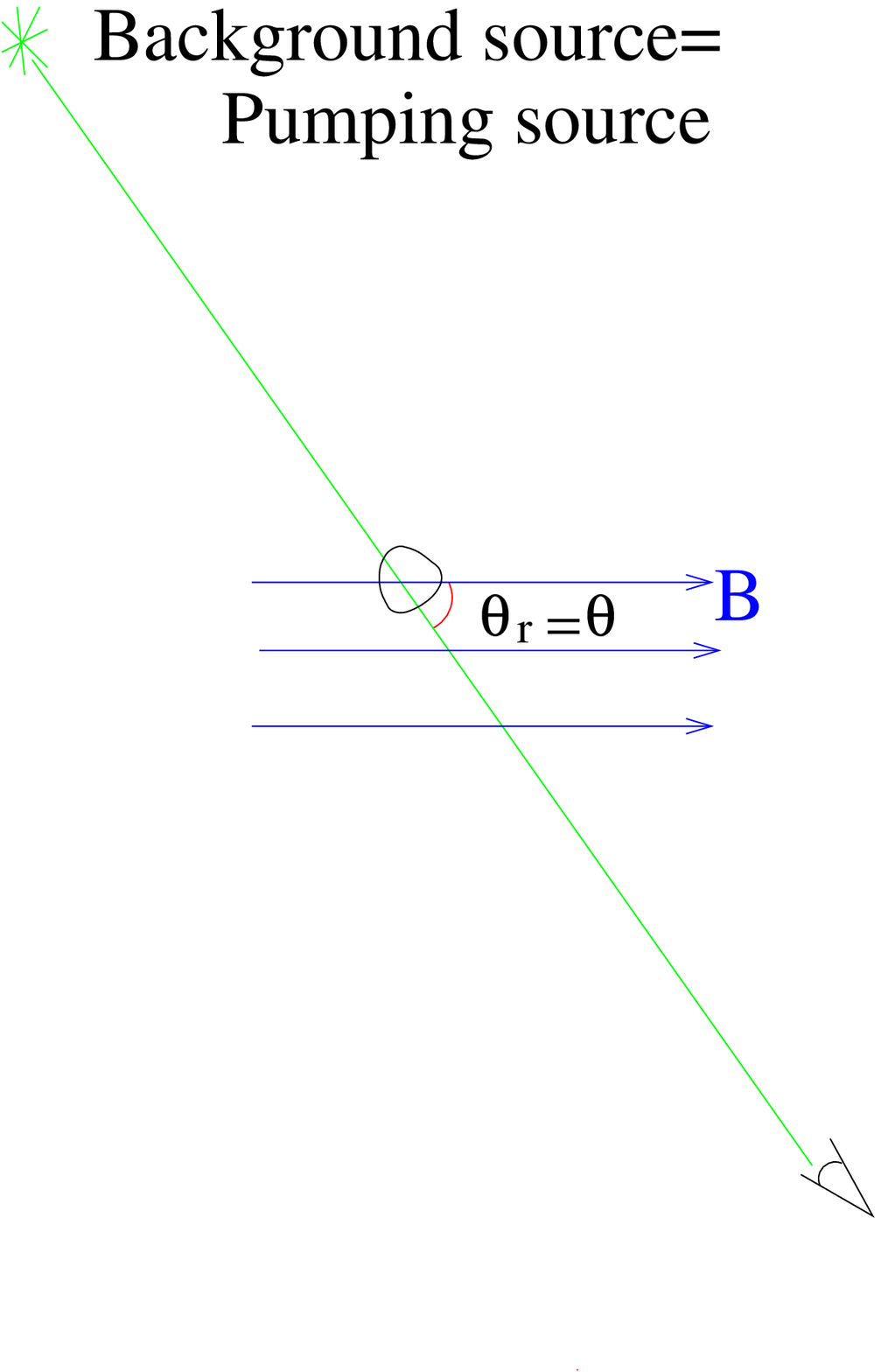}
\caption{Typical astrophysical environment where atomic alignment can happen. A pumping source deposits angular momentum to atoms in the direction of radiation and causes differential occupations on their ground states. In a magnetized medium where the Larmor precession rate $\nu_L$ is larger than the photon arrival rate $\tau_R^{-1}$, however, atoms are realigned with respect to magnetic field. Atomic alignment is then determined by $\theta_r$, the angle between the magnetic field and the pumping source. Observed polarization depends on both $\theta_r$ and $\theta$, the angle between the magnetic field and the line of sight. In general, there are two situations: {\it left}, the alignment is produced by a pumping source while we observe another weak background source whose light passes through the aligned medium; {\it right}, the background source coincides with the pumping source, in this case, $\theta_r=\theta$.}
\label{geometry}
\end{figure}

The basic idea of the atomic alignment is quite
simple. The alignment is caused by
the anisotropic deposition of angular momentum from photons\footnote{We do not consider interactions with other particles (see Fineschi \& Landi Degl'Innocenti 1992).}. In typical
 astrophysical situations the radiation
flux is anisotropic (see Fig.\ref{geometry}). As the photon
spin is along the direction of its propagation, we expect that atoms
scattering the radiation from a light beam to be  aligned. Such an alignment happens in terms of 
the projections of angular momentum
to the direction of the incoming light. For atoms to be aligned,
 their ground state should have non-zero angular momentum. Therefore fine (or hyperfine) structure is necessary to enable various projection of atomic angular momentum to exist in their ground state. 

\subsection{Basics of atomic alignment}
\begin{figure}
\includegraphics[%
  width=0.45\textwidth,
  height=0.3\textheight]{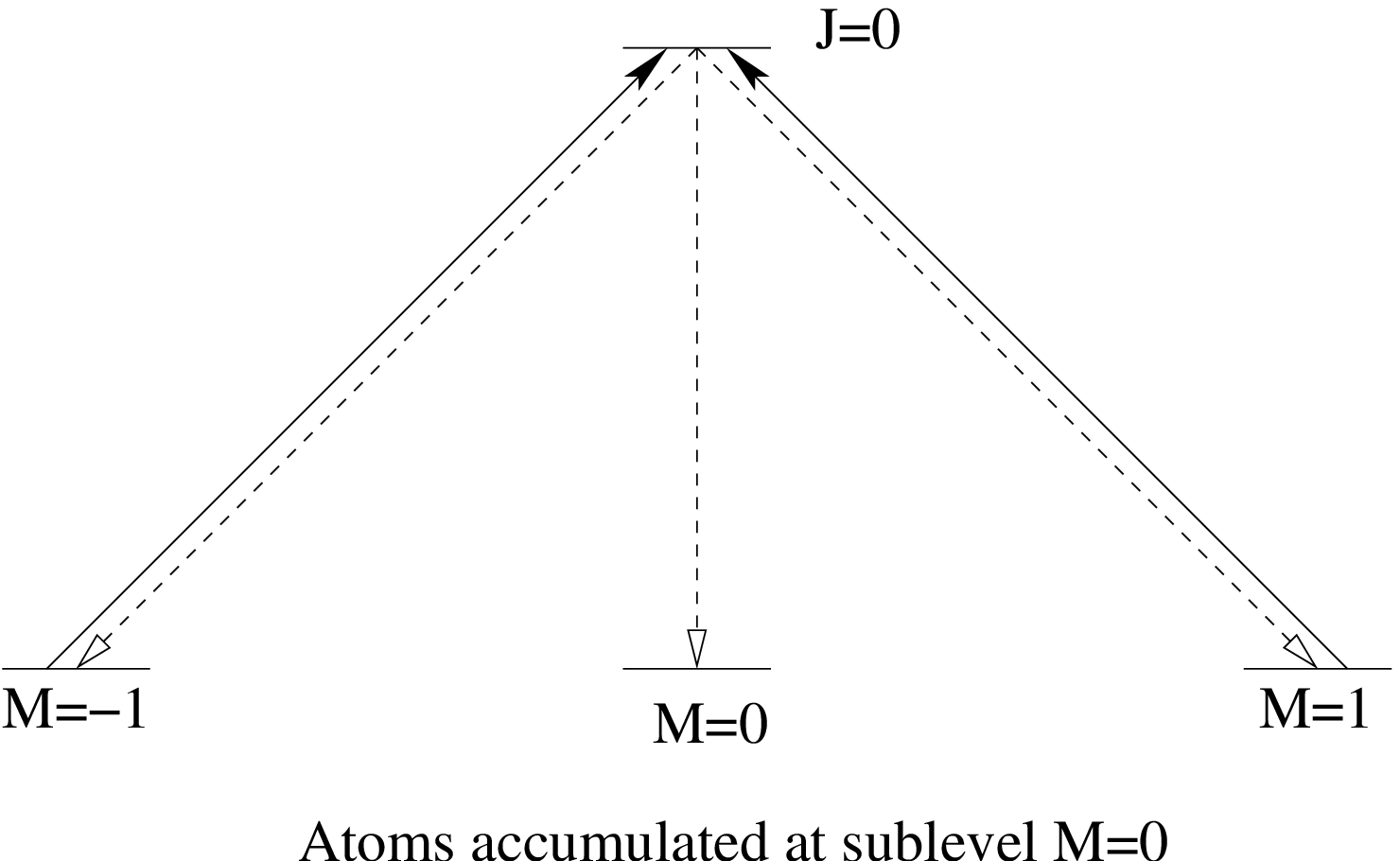}
\includegraphics[%
  width=0.45\textwidth,
  height=0.3\textheight]{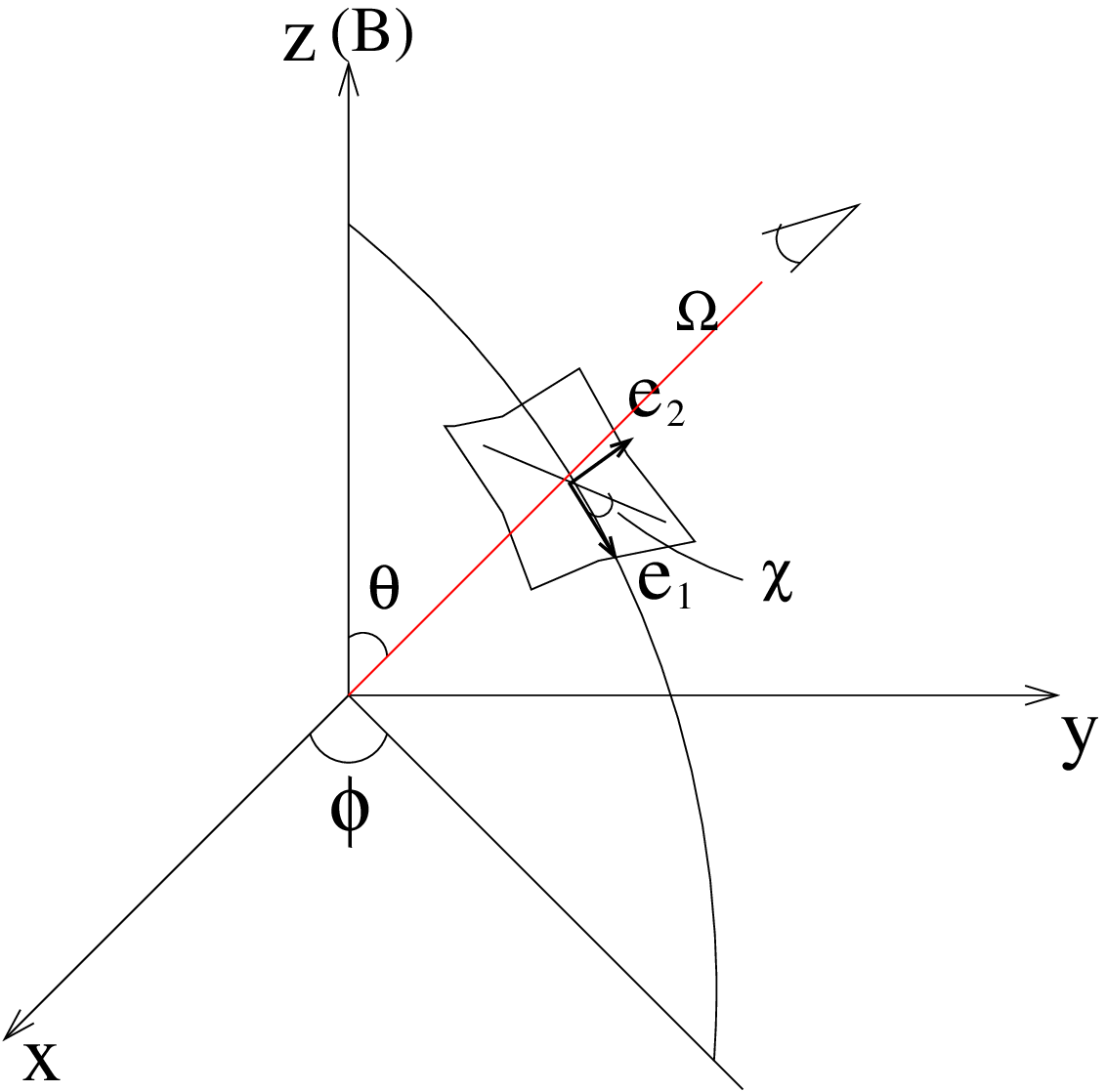}
\caption{{\it Left}: a toy model to illustrate how atoms are aligned by anisotropic light.
Atoms accumulate in the ground sublevel $M=0$ as radiation removes atoms from the ground states $M=1$ and $M=-1$; {\it Right}: Radiation geometry and the polarization vectors in a given coordinate system. ${\bf \Omega}$ is the direction of radiation, $\chi$ is the positional angle of a linear polarization.}
\label{nzplane}
\end{figure}

Let us discuss a toy model that provides an intuitive  insight into  the 
physics of atomic alignment.
Consider an
atom with its ground state corresponding to the total angular momentum
$I=1$ and the upper state corresponding to the angular momentum $I=0$ (Varshalovich 1971).
If the projection of the angular momentum to the direction of the
incident resonance photon beam is $M$, the upper state $M$ can
have values $-1$, $0$, and $1$, while for the upper state M=0 (see Fig.\ref{nzplane}{\it left}). 
The unpolarized
beam contains an equal number of left and right circularly polarized
photons whose projections on the beam direction are 1 and -1. Thus
absorption of the photons will induce transitions from the $M=-1$ and
$M=1$ sublevels. However, the decay from the upper state populates all
the three sublevels on ground state. As the result the atoms accumulate in the $M=0$ ground
sublevel from which no excitations are possible. Accordingly, the optical
properties of the media (e.g. absorption) would change.

The above toy model can also exemplify the
 role of collisions and magnetic field. Without collisions one may expect 
that all atoms
reside eventually at the sublevel of $M=0$. Collisions, however, redistribute
atoms to different sublevels. Nevertheless, as disalignment of the ground
state requires spin flips, it is less efficient than one might naively
imagine (Hawkins 1955). The reduced sensitivity of aligned
atoms to disorienting collisions makes the effect important for various
astrophysical environments.

A magnetic field would mix up different $M$ states. However, it
is clear that the randomization in this situation is not complete
and the residual alignment reflects the magnetic field direction (see Fig.\ref{geometry}). Magnetic mixing happens if the angular
momentum precession rate  is higher than the rate of the
excitation from the ground state, which is true for many astrophysical conditions, e.g., interplanetary medium, ISM, intergalactic medium, etc. 

{\it Since conservation and transfer of angular momentum is essential for the problem we deal with, it is necessary to quantize the radiation field as the atomic states are quantized}. The classical theory can give a qualitative interpretation which shall be utilized in this paper to provide an intuitive picture. However, quantitative results have to be obtained by the quantum-mechanical treatment, namely, density matrix description for both atoms and the radiation. The key issue here is that the focus is on angular momentum instead of energy. As we are interested in exchange of angular momentum between photons and atoms, we have to quantize the radiation field.


All in all, in order to be aligned, first, atoms should have enough 
degrees of freedom: namely, the quantum angular momentum number must be 
$\ge 1$. Second, the incident flux must be anisotropic. 
Moreover, the collisional rate should not be too high. While the latter
requires special laboratory conditions, it is applicable to many astrophysical environments such as the outer
layers of stellar atmospheres, the interplanetary,
interstellar, and intergalactic medium.

\subsection{Relevant time-scales}
\label{relscales}
\begin{figure}
\plotone{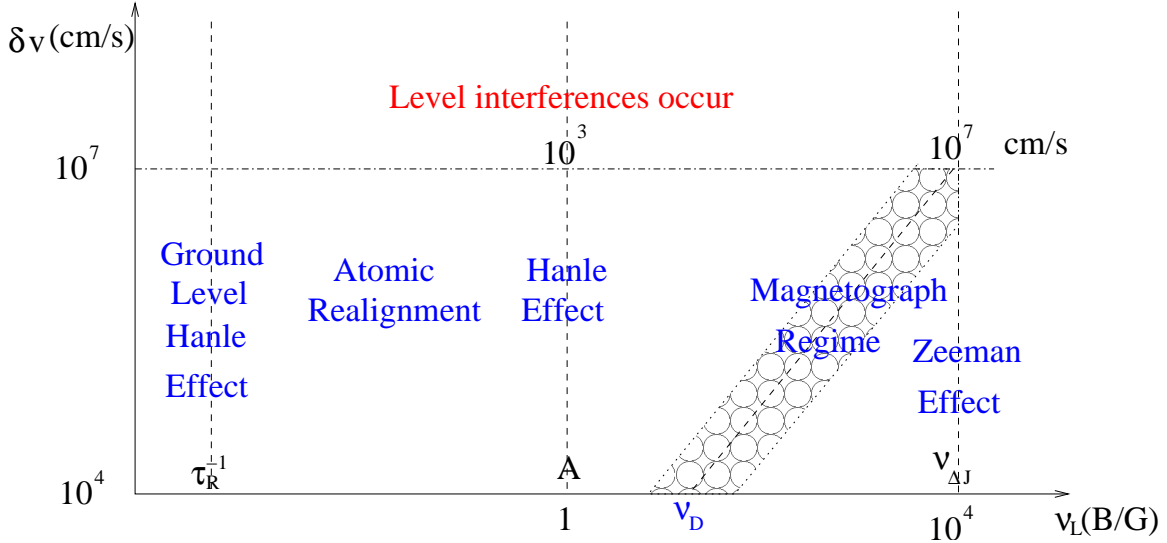}
\caption{Different regimes divided according to the strength of magnetic field and the Doppler line width. Atomic realignment is applicable to weak field ($<1G$) in diffuse medium. Level interferences are negligible unless the medium is substantially turbulent ($\delta v\gtrsim$ 100km/s) and the corresponding Doppler line width becomes comparable to the fine level splitting $\nu_{\Delta J}$. For strong magnetic field, Zeeman effect dominates. When magnetic splitting becomes comparable to the Doppler width, $\sigma$ and $\pi$ components can still distinguish themselves through polarization, this is the magnetograph regime; Hanle effect is dominant if Larmor period is comparable to the lifetime of excited level $\nu_L^{-1}\sim A^{-1}$; similarly, for ground Hanle effect, it requires Larmor splitting to be of the order of photon pumping rate; for weak magnetic field ($<1G$) in diffuse medium, however, atomic alignment is the main effect provided that $\nu_L>\tau_R^{-1}$.}
\label{regimes}
\end{figure}
{\large
\begin{table*}
\begin{tabular}{||c|c|c|c|c||}
\hline
 Atom&
 Lower state&
 Upper state&
 Wavl(\AA)&
 $P_{max} (J_l\rightarrow J_u)$
 \tabularnewline
\hline 
\hline
S II&$4S^o_{3/2}$&
$4P_{1/2,3/2,5/2}$& 1250-1260&12\%($3/2\rightarrow 1/2$)\\
\hline
Cr I&
$a7S_3$&
$7P^o_{2,3,4}$&
3580-3606, 4254-4290&5\%($3\rightarrow 2$)\\
\hline
 C II&
$2P^o_{1/2,3/2}$&
$2S_{1/2}$,$2P_{1/2,3/2}$,$2D_{3/2,5/2}$&
$\sim$ 1036.3-1335.7&15\%($3/2\rightarrow 1/2$)\\
\cline{1-1}
\cline{4-5}
Si II&&&989.9-1533.4&7\%($3/2\rightarrow 1/2$)\\
\hline
 O I&
$3P_{2,1,0}$&$3S_1$, $3D_{1,2,3}$&
911-1302.2&29\%($2\rightarrow 2$)\\
\cline{1-1}\cline{3-5}
S I&&$3S_1$,$3P^o_{0,1,2}$, $3D_{1,2,3}$&1205-1826&22\%($1\rightarrow 0$)\\
\hline
 C I&
$3P_{0,1,2}$&
$3P^o_{0,1,2}$,$3D^o_{1,2,3}$&
 1115-1657&18\%($1\rightarrow 0$)\\
\cline{1-1}\cline{4-5}
Si I&&&1695-2529&20\%($2\rightarrow 1$)\\
\cline{1-1}\cline{4-5}
S III &&&$\sim$1012-1202&24.5\%($2\rightarrow 1$)\\
\hline
\hline
\end{tabular}
\caption{Selected alignable atomic species and corresponding transitions. Note only lines above $912 \AA$ are listed. Data are taken from the Atomic Line List http://www.pa.uky.edu/$\sim$peter/atomic/ and the NIST Atomic Spectra Database. The last column gives the maximum polarizations and its corresponding transitions. For those species with multiple lower levels, the polarizations are calculated for shell star ($T_{eff}=15,000$K) in the strong pumping regime; in the weak pumping regime, the maximum polarizations are 19\% for OI transition ($2\rightarrow 2$), and 9\% for SI transition ($2\rightarrow 2$).}
\label{ch3t1}
\end{table*}
}
 
\begin{table*}
\begin{tabular}{||c|c|c|c|c|c||}
\hline
\hline
 Atom&
 Ground state (G)&
 Metastable state (M)&
 Upper state (U)&
 Wavl(\AA) ($G\rightarrow U$)&
 Wavl(\AA) ($M\rightarrow U$)
 \tabularnewline
\hline
\hline
Cr II &$a6S_{5/2}$ & $a6D_{1/2,3/2,...9/2}$&
$z6P^o_{3/2,5/2,7/2}$&
2056-2066& 2741-2767\\
\hline
Ti II&$a4F_{3/2,5/2,7/2,9/2}$&$b4F_{3/2,5/2,7/2,9/2}$&$z4D^o_{1/2,3/2,5/2}$&3058.3-3089&3153-3169\\
\cline{4-6}
&&&$z4F^o_{3/2,5/2,7/2,9/2}$&3218-3255&3310-3348\\
\cline{4-6}
&&&$z4G^o_{5/2,7/2,9/2}$&3362-3411&3445-3501\\
\hline
Ti III& $3F_{2,3,4}$&$3P_{0,1,2}$&$3D^o_{1,2,3}$&1282-1299&1499-1506\\
\cline{3-4}\cline{6-6}
&&$3D_{1,2,3}$&$3D^o_{1,2,3}, 3F^o_{2,3,4}$&&2529-2581\\
\hline
Fe II & $a6D_{9/2,7/2,...,1/2}$&$a6S_{5/2}$&$z6P_{7/2,5/2,3/2}$&2327-2381&4925-5170\\
\cline{4-5}
&&&$z6D_{9/2,7/2,...,1/2},z6F_{11/2,9/2,...,1/2}$&2374-2631&\\
\hline
Fe III&$5D_{4,3,2,1}$&$5S_2$&$5P^o_{3,2,1}$&1123-1132&2062-2079.7\\
\hline
Ca II & $2S_{1/2}$&$2D_{3/2,5/2}$&$2P^o_{1/2,3/2}$&3934-3669 & 8498-8662 \\
\hline
\hline
\end{tabular}
\caption{Examples of absorptions from metastable states. Calculation for Cr II shows that the absorption from metastable level $a6D_{9/2}\rightarrow z6P^o_{9/2}$ has a maximum polarizaton of 21\%.}
\label{ch3t2}
\end{table*}

Various species with fine structure 
can be aligned. A number of selected transitions  that can be used
for studies of magnetic fields are listed in Table \ref{ch3t1}. Why and how are these lines chosen?  First, we gathered all of the prominent interstellar and intergalactic lines (Morton 1975, Savage et al. 2005), from which we take only alignable lines, namely, lines with ground angular momentum number $J_g\geq1$. Then we rule out those species with nuclear spin since hyperfine lines will be dealt with in our companion paper. The
number of prospective transitions increases considerably if we add
QSO lines. In fact, many of the species listed in the Table 1 in Verner, Barthel, \& Tytler (1994).
are alignable and observable from the ground because of the cosmological
redshifts. In this paper we do not deal with such transitions.

In terms of practical magnetic field studies, the variety of available species is important in many aspects. One of them is a possibility of
getting additional information about environments. Let us illustrate this by considering
the various rates (see Table \ref{ch3t2}) involved. Those 
are 1) the rate of the Larmor precession, $\nu_L$, 2) the rate of the optical pumping, $\tau_R^{-1}$, 3) the rate of collisional randomization, $\tau_c^{-1}$,
4) the rate of the transition within ground state, $\tau^{-1}_T$.  
In many cases $\nu_L>\tau_R^{-1}>\tau_c^{-1}, \tau_T^{-1}$. 
Other relations are possible, however. If $\tau_T^{-1}>\tau_R^{-1}$, the transitions within the sublevels of ground state need to be taken into account and relative distribution among them will be modified (see \S\ref{weakp}). Since emission is spherically symmetric, the angular momentum in the atomic system is preserved and thus alignment persists in this case. In the case $\nu_L<\tau_R^{-1}$, the magnetic field does not affect the atomic occupations and atoms are aligned with respect to the direction of radiation (see Fig.\ref{geometry}). From the expressions in Table~\ref{difftime}, we see, for instance, that magnetic field can realign CII only at a distance $r\gtrsim7.7$Au from an O star if the magnetic field strength $\sim 5\mu$G.

If the Larmor precession rate $\nu_L$ is comparable to any of the other rates,
the atomic line polarization becomes sensitive to the strength of the magnetic field. In these situations, it is possible to get information about the {\it magnitude} of magnetic
field. In this paper we consider the regime where $\nu_L$ is much smaller that the decay rate of the excited state, which means that we disregard the Hanle effect.
{\large
\begin{table}
\begin{tabular}{||c|c|c|c||}
\hline
\hline
$\nu_L$(s$^{-1}$)&$\tau_R^{-1}$(s$^{-1}$)&$\tau_T^{-1}$(s$^{-1}$)&$\tau_c^{-1}$(s$^{-1}$)\\
\hline
$\frac{eB}{m_ec}$&$B_{J_lJ_u}I$&$A_m$&max($f_{kj},f_{sf}$) \\
\hline
$88(B/5\mu$ G)& $7.4\times 10^{5}\left(\frac{R_*}{r}\right)^2$&2.3$\times 10^{-6}$&$6.4\left(\frac{n_e}{0.1{\rm cm}^{-3}}\sqrt{\frac{8000{\rm K}}{T}}\right)\times 10^{-9}$\\
\hline
\hline
\end{tabular}
\caption{Relevant rates for atomic alignment. $A_m$ is the magnetic dipole emission rate for transitions among J levels of the ground state of an atom. $f_{kj}$ is the inelastic collisional transition rates within ground state due to collisions with electrons or hydrogens, and $f_{sp}$ is the spin flip rate due to Van der Waals collisions (see \S\ref{collision}). In the last row, example values for C II are given. $\tau_R^{-1}$ is calculated for an O type star, where $R_*$ is the radius of the star radius and r is the distance to the star.}
\label{difftime}
\end{table}
}
Fig.\ref{regimes} illustrates the regime of magnetic field strength where atomic realignment applies. Atoms are aligned by the anisotropic radiation at a rate of $\tau_R^{-1}$. Magnetic precession will realign the atoms in their ground state if the Larmor precession rate $\nu_L>\tau_R^{-1}$. In contrast, if the magnetic field gets stronger so that Larmor frequency becomes comparable to the line-width of the upper level, the upper level occupation, especially coherence is modified directly by magnetic field, this is the domain of Hanle effect, which has been extensively discussed for studies of solar magnetic field (see Landi Degl'Innocenti 2004 and references therein). When the magnetic splitting becomes comparable to the Doppler line width $\nu_D$, polarization appears, this is the ``magnetograph regime" (Landi Degl'Innocenti 1983b). For magnetic splitting $\nu_L\gg \nu_D$, the energy separation is enough to be resolved, and the magnetic field can be deduced directly from line splitting in this case. If the medium is strongly turbulent with $\delta v\sim 100$km/s (so that the Doppler line width is comparable to the level separations $\nu_D\sim \nu_{\Delta J}$), interferences occur among these levels and should be taken into account.

Long-lived alignable metastable states that are present for
some atomic species between upper and lower states may act as
proxies of ground states.  Absorptions from these metastable levels
 can also be used as diagnostics for magnetic field therefore.
We present a few selected species with such states in Table~\ref{ch3t2}, but
further in the present paper deal only with one representative of this
class, namely, Cr II. The life time of the metastable level is an
important timescale. However, in the formalism of the present paper one can account for it
via modifying the effective 
$\tau_T$ to account for the decay of the metastable states.

\subsection{Ground state alignment and absorptions}
\begin{figure}
\plottwo{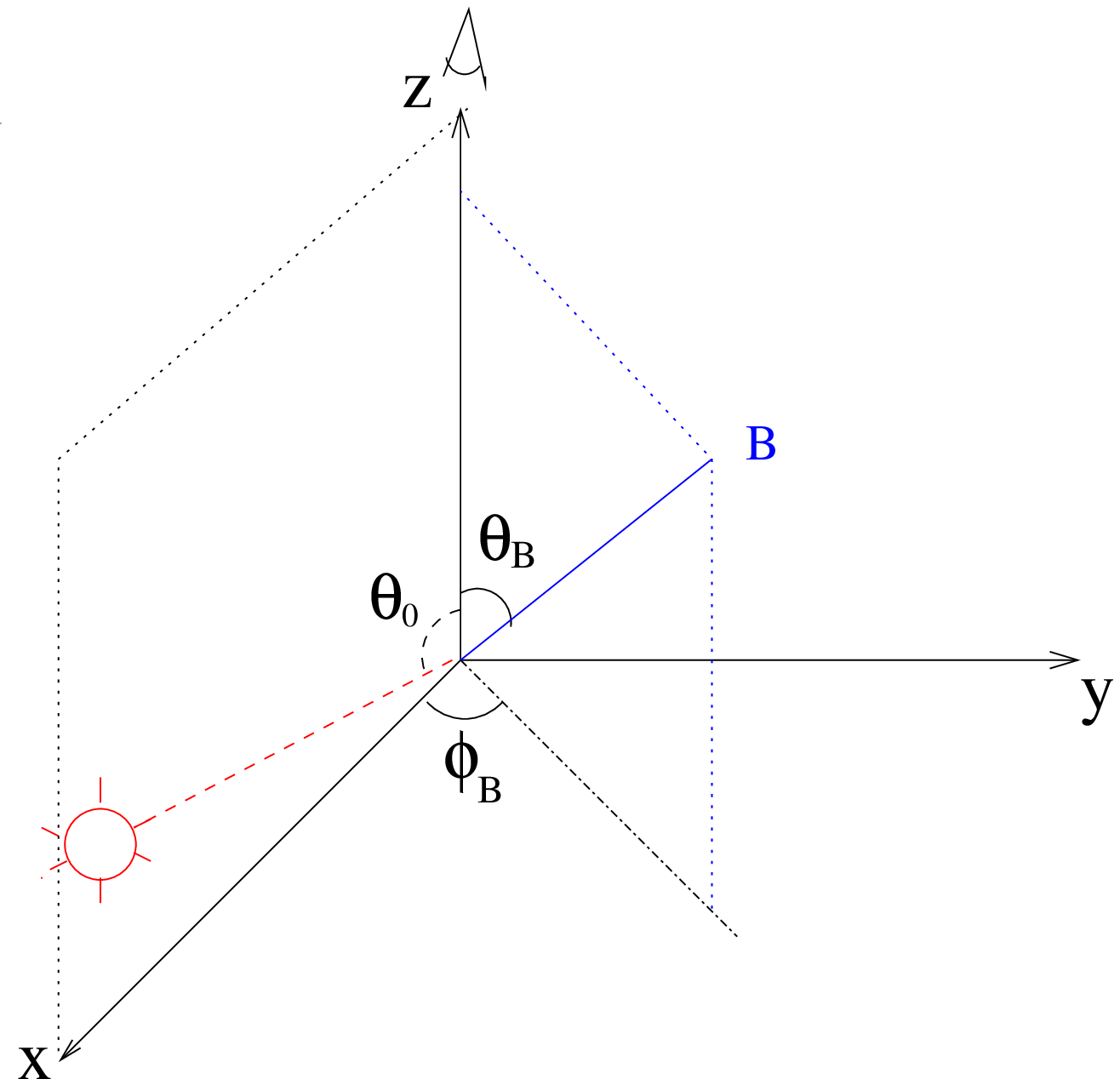}{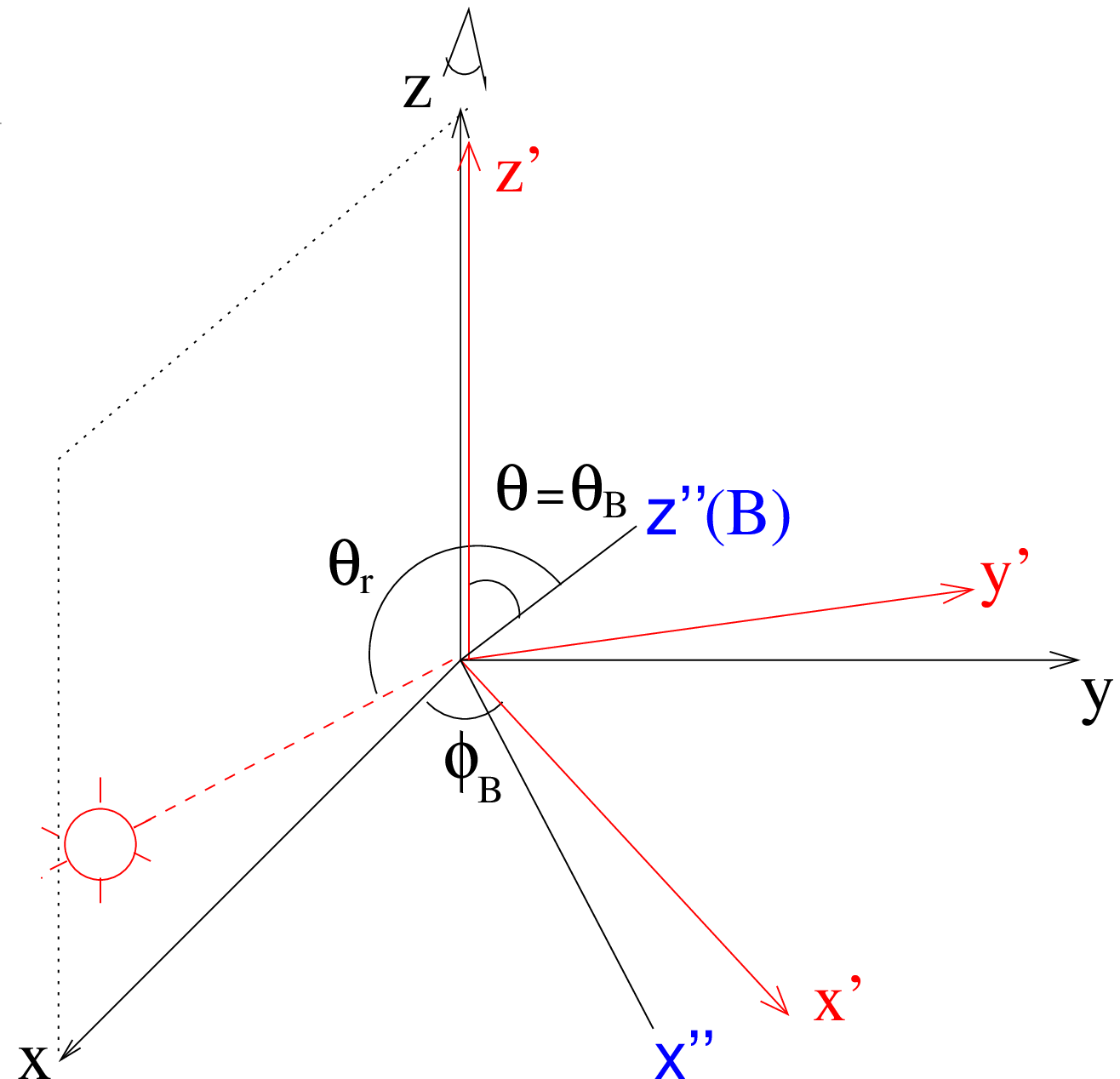}
\caption{{\it Left}: geometry of the observational frame. In this frame, the line of sight is z axis, together with the incident light, they specify the x-z plane. Magnetic field is in ($\theta_B, \phi_B$) direction; {\it Right}: transformation to the ``theoretical frame" where magnetic field defines z" axis. This can be done by two successive rotations specified by Euler angles ($\phi_B, \theta_r$) (see App.\ref{euler} for details). The first rotation is from xyz coordinate system to x'y'z' coordinate system by an angle $\phi_B$ about the z-axis, the second is from x'y'z' coordinate system to x"y"z" coordinate system by an angle $\theta$  about the y'-axis. Atomic alignment and transitions are treated in the ``theoretical" frame where the line of sight is in ($\theta, \pi$) direction and the incident radiation is in ($\theta_r,\phi_r$) direction. }
\label{radiageometry}
\end{figure}
The geometry of the radiation system is illustrated by Fig.\ref{radiageometry}{\it left}. The origin of this frame is defined as the location of the atomic cloud. The line of sight defines z axis, and together with direction of radiation, they specify x-z plane. The x-y plane is thus the plane of sky. In this frame, the incident radiation is coming from  ($\theta_0, 0$), and the magnetic field is in the  direction ($\theta_B, \phi_B$). 

The magnetic field is chosen as the quantization axis ($z"$) for the atoms. Alignment shall be treated in the frame $x"y"z"$ (see App.\ref{euler} how these two frames are related). In this ``theoretical'' frame, the line of sight is in ($\theta, \pi$) direction (i.e., the x"-z" plane is defined by the magnetic field and the line of sight, see Fig.\ref{radiageometry}{\it right}), and the radiation source is directed along ($\theta_r, \phi_r$).

Absorption is directly associated with the alignment of the ground state and the direction of observation (Fig.\ref{geometry}).  For absorption, we shall demonstrate later that the Stokes parameter (see App.\ref{Stokes}) $U=0$ if the background is unpolarized. This means that the linear polarization is either parallel or perpendicular to the magnetic field. This happens because of fast precession of atoms in the ground state. Direction of polarization thus traces $\phi_B$ with a $90^o$ degeneracy. For absorptions in the optically thin case\footnote{In solar case, one has to deal with radiative transfer. For diffuse medium, however, this assumption is reasonable.}, the Stokes parameters $Q=-\eta_1dI_0, I=I_0(1-\eta_0d)$, where $\eta_i$ are absorption coefficients and given by Eq.\ref{Mueller}. d is the thickness of the medium. $I_0$ is the intensity of background source. Thus the degree of polarization $P$ per unit optical depth $\tau=\eta_0d$ is given by

\be
\frac{P}{\tau}=\frac{Q}{I\eta_0d}\simeq\frac{-\eta_1}{\eta_0}=\frac{1.5\sigma^2_0(J_l)\sin^2\theta w^2_{J_lJ_u}}{\sqrt{2}+\sigma^2_0(J_l)(1-1.5\sin^2\theta)w^2_{J_lJ_u}}.
\label{absorb}
\ee
$\sigma^2_0$ depends on $\theta_r$ and is the normalized dipole component of the ground state density matrix, which is the main parameter we calculate in this paper. $w^2_{J_lJ_u}$ is defined by Eq.(\ref{w2}) and the corresponding numerical values are given in Table~\ref{ch3t2}. For a generic case, where the background source (e.g. QSO) is polarized and optical depth is finite\footnote{We neglect emission here.}, we can get in the first order approximation
\bea
I&=&(I_0+Q_0)e^{-\tau(1+\eta_1/\eta_0)}+(I_0-Q_0)e^{-\tau(1-\eta_1/\eta_0)},\nonumber\\
Q&=&(I_0+Q_0)e^{-\tau(1+\eta_1/\eta_0)}-(I_0-Q_0)e^{-\tau(1-\eta_1/\eta_0)},\nonumber\\
U&=&U_0e^{-\tau}, V=V_0e^{-\tau},
\label{generalabsorb}
\eea
in which ${I_0, Q_0, U_0, V_0}$ are the Stokes parameters of background source. Optical depth is also modified by the alignment. The line ratio of a doublet varies as
\be
\frac{\tau_1}{\tau_2}=\frac{B_{J_lJ_1}[\sqrt{2}+\sigma^2_0(J_l)(1-1.5\sin^2\theta)w^2_{J_lJ_{2}}]}{B_{J_lJ_2}[\sqrt{2}+\sigma^2_0(J_l)(1-1.5\sin^2\theta)w^2_{J_lJ_{1}}]},
\label{tauratio}
\ee
where $B_{J_lJ_u}$ represents Einstein coefficient. 
The next section will be devoted to explaining how the atomic alignment ($\sigma^2_0$) can be calculated using QED theory. Readers who are not interested in the technical details, however, can skip that section and go directly to the results of calculations for the species in Table 1 that we present in \S\ref{mainresults}. The value of \S\ref{theory} is that it provides the approach for calculating the alignment of atoms other than those discussed in the present paper.

\section{Alignment theory}
\label{theory}

In what follows, we consider a realistic atomic system, which can have multiple upper levels $J_u,M_u$ and lower levels $J_l,M_l$.   When such an atomic system interacts with resonant radiation, there will be a photoexcitation followed by spontaneous emission. The atomic occupation is thus determined by the balance of the three processes: absorption, emission and magnetic precession among the sublevels of the state. The radiative transition rates can be constructed from the transition probabilities given by Eq.(\ref{smalla}). Then additional transformation should be made to express the density matrices in irreducible form (see App.\ref{irredmatrix}). The actual derivation of the equations that govern the statistical equilibrium of the atomic occupations is rather complicated and can be found in Landi Degl’Innocenti \& Landolﬁ (2004). Below are the statistical equilibrium equations for the upper and ground state (Degl’Innocenti \& Landolﬁ 2004): 
\bea
\dot{\rho^k_q}(J_u)&+&i 2\pi\nu_Lg_uq\rho^k_q(J_u) = -\sum_{J_l}A(J_u\rightarrow J_l)\rho^k_q(J_u)+\sum_{J_lKQk'q'}[J_l]B(J_l\rightarrow J_u)(-1)^{k'+q'}\nonumber\\
&&(3[k,k',K])^{1/2}\left\{\begin{array}{ccc} 
1 & J_u & J_l\\1& J_u & J_l\\ K &k& k' \end{array}\right\}\left(\begin{array}{ccc}
k & k' & K\\ q & q' & Q \end{array}\right)\bar{J}^K_Q \rho^{k'}_{-q'}(J_l)
\label{evolution}
\eea
\bea
\dot{\rho^k_q}(J_l)&+&i 2\pi\nu_Lg_lq\rho^k_q(J_l) = \sum_{J_u}(-1)^{J_l+J_u+k+1}[J_u]\left\{\begin{array}{ccc}
J_l & J_l & k\\J_u & J_u &1\end{array}\right\}A(J_u\rightarrow J_l)\rho^k_q(J_u) \nonumber \\
&-&[J_l]\sum_{J_uKQk'q'}B(J_l\rightarrow J_u)(-1)^{J_l-J_u+k+k'+q'+1}(3[k,k',K])^{1/2}\nonumber \\
&&\left\{\begin{array}{ccc} 
1 & 1 & K\\J_l& J_l & J_u\end{array}\right\}\left\{\begin{array}{ccc} 
k & k' & K\\J_l& J_l & J_l\end{array}\right\}\left(\begin{array}{ccc}
k & k' & K\\ q & q'& Q\end{array}\right)\bar{J}^K_Q \rho^{k'}_{-q'}(J_l),
\label{evolutiong}
\eea
where $\rho^k_q$ and ${\bar J}^K_Q$ are the irreducible form of density matrix of atoms and radiation, respectively. The evolution of upper state ($\rho^k_q(J_u)$) is represented by Eq.(\ref{evolution}) and the ground state ($\rho^k_q(J_l)$) is described by Eq.(\ref{evolutiong}). The second terms on the left side of Eq.(\ref{evolution},\ref{evolutiong}) represent mixing by magnetic field, where $g_u$ and $g_l$ are the Land\'e factors for the upper and ground level. For the upper level, this term can be ignored as we consider a regime where the magnetic field is much weaker than Hanle field\footnote{For the Hanle effect to be dominant, magnetic splitting ought to be comparable to the energy width of the excited level; see Fig.\ref{regimes}.} ($\nu_L\ll A/g_u$). The two terms on the right side of Eq.(\ref{evolution}, \ref{evolutiong}) are due to spontaneous emissions and the excitations from ground level. Transitions to all upper states are taken into account by summing over $J_u$ in Eq.(\ref{evolutiong}). Vice versa, for an upper level, transitions to all ground sublevels ($J_l$) are summed up in Eq.(\ref{evolution}). Let us remind our reader, this multi-level treatment is well justified as level crossing interferences do not exist in the weak field regime unless the medium gets strongly turbulent so that Doppler width becomes comparable to the energy separation between levels ($\delta v\gtrsim 100$km/s, see Fig.\ref{regimes}). The matrix with big ``\{ \}" represents the 6-j or 9-j symbol, depending on the size of the matrix. When more than two angular momentum vectors are coupled, there is more than one way to add them up and form the same resultant. The 6-j (or 9-j) symbol appears in this case as a recoupling coefficient describing transformations between different coupling schemes (see App.\ref{angmtmcpl}). Throughout this paper, we define $[j]\equiv 2j+1$, which means $[J_l]=2J_l+1$ and $[J_u]=2J_u+1$. 
The Einstein spontaneous emission rate is defined as the total probability of an atom from a specific state $J',M'$ to all states of $J$: $A=\sum_{M}a$. The Einstein coefficients for stimulated transitions B are related to $A$ by: 
\be
[J_u]A(J_u\rightarrow J_l)=2h\nu^3/c^2[J_l]B(J_l\rightarrow J_u)=2 h\nu^3/c^2[J_u]B(J_u\rightarrow J_l),
\label{Ein}
\ee
where $[J]$ is the statistical weight of the initial level $J$. 
 Stimulated emission is neglected since $A/g_u\gg\nu_L>\tau_R^{-1}\sim B(J_u\rightarrow J_l) {\bar J}^0_0$. Note for spontaneous emission and magnetic mixing, $k,q$ are conserved because of the symmetric nature of these processes. Indeed, magnetic mixing and emission are axially symmetric processes.  

The excitation depends on  
\be
\bar{J}^K_Q=\int d\nu \frac{\nu_0^2}{\nu^2}\xi(\nu-\nu_0)\oint \frac{d\Omega}{4\pi}\sum_{i=0}^3{\cal J}^K_Q(i,\Omega)S_i(\nu, \Omega),
\label{incidentrad}
\ee
which is the radiation tensor of the incoming light averaged over the whole solid-angle and the 
line profile $\xi(\nu-\nu_0)$. $S_i=[I,\, Q,\, U,\, V]$ represent the Stokes parameters (App.\ref{Stokes}). In many cases, the
radiation source is far enough
 that it can be treated as a point source. We consider here unpolarized incident light, thus $Q=U=V=0$. The irreducible unit tensors for Stokes parameters $I,Q,U$ (App.\ref{Stokes}) are:
\bea
{\cal J}^0_{0}(i,\Omega)&=&\left(\begin{array}{l}1\\0\\0\end{array}\right),\;~\;\; {\cal J}^2_{0}(i,\Omega)=\sqrt {2}\left[\begin{array}{l} ( 1-1.5\sin^2\theta )/2\\-3/4 \sin^2\theta\\0\end{array}\right],\nonumber\\
 {\cal J}^2_{\pm2}(i,\Omega)&=&\sqrt{3}\left[\begin{array}{l}\sin^2\theta/4\\-( 1+\cos^2\theta )/4\\\mp i\cos \theta/2  \end{array}\right]e^{\pm 2i\phi},\;~\;\;
{\cal J}^2_{\pm1}(i,\Omega)=\sqrt{3}\left(\begin{array}{l}\mp\sin 2 \theta/4 \\\mp\sin2 \theta/4 \\-i\sin \theta/2 
\end{array}\right)e^{\pm i\phi}.
\label{irredrad}
\eea
They are determined by its direction $\Omega$ and the reference chosen to measure the polarization (Fig.\ref{nzplane}{\it right}). As addressed earlier, the line of sight is in $\Omega=(\theta, \pi)$ in the ``theoretical frame". For the incoming radiation from ($\theta_r,\phi_r$), according to Eq.(\ref{irredrad}), the nonzero elements of the radiation tensor are:

\be
\bar{J}^0_0=I_*, \bar{J}^2_0=\frac{W_a}{2\sqrt{2}W}(2-3\sin^2\theta_r)I_*, \bar{J}^2_{\pm2}=\frac{\sqrt{3}W_a}{4W}\sin^2\theta_rI_*{e^{\pm 2i\phi_r}}, \bar{J}^2_{\pm1}=\mp\frac{\sqrt{3}W_a}{4W}\sin2\theta_r I_*{e^{\pm i\phi_r}} 
\label{irredradia}
\ee
where W is the dilution factor of the radiation field, which can be divided into anisotropic part $W_a$ and isotropic part $W_i$ (Bommier \& Sahal-Brechot 1978). In the case of a point source, $W_i=0$. If $W_i\neq 0$, the degree of alignment and polarization will be reduced. The solid-angle averaged intensity for a black-body radiation source is 
\be
I_*=W\frac{2h\nu^3}{c^2}\frac{1}{e^{h\nu/k_BT}-1}.
\label{Idilu}
\ee
By setting the first terms on the left side of Eq.(\ref{evolution},\ref{evolutiong}) to zeros, we can obtain the following linear equations for the steady state occupations in the ground states of atoms: 
\bea
&2\pi i\rho^k_q(J_l)q g_l\nu_L&-\sum_{J_u}\frac{A(J_u\rightarrow J_l)}{\sum_{J'_l} A(J_u\rightarrow J'_l)}\sum_{J'_l}[J'_l]B(J'_l\rightarrow J_u) \sum_{k'q'}\rho^{k'}_{-q'}(J'_l)\sum_{KQ}(-1)^{k+k'+q'}\nonumber \\
&&(3[k,k',K])^{1/2}\left(\begin{array}{ccc}
k & k' & K\\ q & q'& Q\end{array}\right)\bar{J}^K_Q\left[(-1)^{J_l+J_u}[J_u]\left\{\begin{array}{ccc}
J_l & J_l & k\\J_u & J_u &1\end{array}\right\}\left\{\begin{array}{ccc} 
1 & J_u & J'_l\\1& J_u & J'_l\\ K &k& k' \end{array}\right\}-\right. \nonumber \\
& &\left.(-1)^{J_l-J_u}\delta_{J_lJ'_l}\left\{\begin{array}{ccc} 
1 & 1 & K\\J_l& J_l & J_u\end{array}\right\}\left\{\begin{array}{ccc} 
k & k' & K\\J_l& J_l & J_l\end{array}\right\}\right]=0
\label{lowlevel}
\eea
The multiplet effect is counted for by summing over $J_u$. The alignment of atoms induces polarization for absorption lines even though the background radiation is unpolarized. The corresponding absorption coefficients $\eta_i$ are (Landi Degl'Innocenti 1984)
\be
\eta_i(\nu, \Omega)=\frac{h\nu_0}{4\pi}Bn(J_l)\xi(\nu-\nu_0)\sum_{KQ}(-1)^Kw^K_{J_lJ_u}\sigma^K_Q(J_l ){\cal J}^K_Q(i, \Omega),
\label{Mueller0}
\ee
where $n(J_l)=n\sqrt{[J_l]}\rho_0^0(J_l)$ is the total atomic population on level $J_l$, $\sigma^K_Q\equiv\rho^K_Q/\rho^0_0$, and
\be
w^K_{J_lJ_u}\equiv\left\{\begin{array}{ccc}1 & 1 & K\\J_l&J_l& J_u\end{array}\right\}/\left\{\begin{array}{ccc}1 & 1 & 0\\J_l&J_l& J_u\end{array}\right\}.
\label{w2}
\ee
The values of $w^2_{JJ'}$ for different pairs of $J,J'$ are listed in Table~\ref{ch3t3}.
{\large
\begin{table}
\begin{tabular}{||c|ccc|ccc|ccc||}
\hline
\hline
 $J$&
 \multicolumn{3}{c|}{1}&
 \multicolumn{3}{c|}{3/2}&
 \multicolumn{3}{c||}{2}
  \tabularnewline
\hline 
 $J'$&
 0&1&2& 1/2&3/2&5/2 &1&2&3\\
\hline
 $w^2_{JJ'}$&
 1&-0.5&0.1& 0.7071&-0.5657&0.1414 &0.5916&-0.5916&0.1690\\
 \hline
 \hline 
 $J$&
 \multicolumn{3}{c|}{5/2}&
 \multicolumn{3}{c|}{3}&
 \multicolumn{3}{c||}{7/2}
  \tabularnewline
\hline 
 $J'$&
 3/2&5/2&7/2 &2&3&4& 5/2&7/2&9/2\\
\hline
 $w^2_{JJ'}$&
 0.5292&-0.6047&0.1890& 0.4899&-0.6124&0.2041& 0.4629&-0.6172&0.2160\\
 \hline
 \hline
 $J$&
 \multicolumn{3}{c|}{4}&
 \multicolumn{3}{c|}{9/2}&
 \multicolumn{3}{c||}{5}
  \tabularnewline
\hline 
 $J'$&
 3&4&5& 7/2&9/2&11/2 &4&5&6\\
\hline
 $w^2_{JJ'}$&
 0.4432&-0.6205&0.2256& 0.4282&-0.6228&0.2335& 0.4163&-0.6245&0.2402\\
 \hline
\hline
\end{tabular}
\caption{Numerical values of $w^2_{JJ'}$. $J$ is the J value of the initial level and $J'$ is that of the final level.}
\label{ch3t3}
\end{table}
}

 Since there is no magnetic coherence because of fast magnetic mixing on ground level, $\sigma^K_{q\neq 0}=0$,
\be
\eta_i(\nu, \Omega)=\frac{h\nu_0}{4\pi}Bn(J_l)\xi(\nu-\nu_0)\sum_{K}(-1)^K w^K_{J_lJ_u}\sigma^K_0(J_l ){\cal J}^K_0(i, \Omega),
\label{Mueller}
\ee
From the above equation, we see that $\eta_2=0$ as ${\cal J}^k_0(2,\Omega)=0$ (Eq.\ref{irredrad}) for our choice of the reference direction of polarization. In this case $\chi$ is either $0^o$ or $90^o$, which means the polarization is either parallel or perpendicular to the projection of magnetic field in the plane of sky (Fig.\ref{nzplane}{\it right}).

Comparing with absorption, emission is negligible as emitted intensity $I_{em}\propto \rho_u\propto {\bar J}^0_0$ (the solid-angle averaged intensity) which is reduced by a dilution factor $W\ll 1$ (Eq.\ref{Idilu}). 

Thereby for optically thin case, the Stokes parameters of an absorbed unpolarized radiation are,
\be
\begin{array}{cc}
I=(1-\eta_0d)I_0,& Q=-\eta_1dI_0
\end{array}
\label{medabs}
\ee
If we drop the assumption of unpolarized background source, the more general expression for finite optical depth would be  
\be
I=(I_0+Q_0)e^{-(\eta_0+\eta_1)d}+(I_0-Q_0)e^{-(\eta_0-\eta_1)d},\,
Q=(I_0+Q_0)e^{-(\eta_0+\eta_1)d}-(I_0-Q_0)e^{-(\eta_0-\eta_1)d},\nonumber\\
U=U_0e^{-\tau},\,V=V_0e^{-\tau}
\ee
where $I_0, Q_0, U_0$ are the Stokes parameters of the background radiation, which can be from a weak background source or the pumping source itself. $d$ refers to the thickness of medium.

For unpolarized pumping light, the radiation has only components with $K=0, 2$. As the result, the density tensor of the atom has only even k components, while odd k components will vanish according to the selection rules of ``3j", ``6j", ``9j" symbols (see App.~\ref{angmtmcpl}). Physically, this means atoms are aligned but not oriented.

\section{Alignment of various atomic species}
\label{mainresults}
\subsection{Single ground level species: S II, \& Cr I}
\label{Cr II}

The S II and Cr I absorption lines are observed in interstellar medium (Morton 1975). We shall discuss S II first. S II has a ground level $4S^o_{3/2}$ and upper levels $4P_{1/2,3/2,5/2}$, which means $J_l=3/2$, $J_u=1/2,3/2,5/2$. According to the triangle rule (App.\ref{angmtmcpl}) of the 3j symbol in Eq.(\ref{irreducerho}), the irreducible density tensor of the ground state $\rho^k_q(J_l)$ has components with $k=3/2-3/2, 3/2-1/2,...,3/2+3/2=0,1,2,3$. For unpolarized pumping light, we only need to consider even components $k=0,2$ as explained earlier.

\begin{figure}
\plottwo{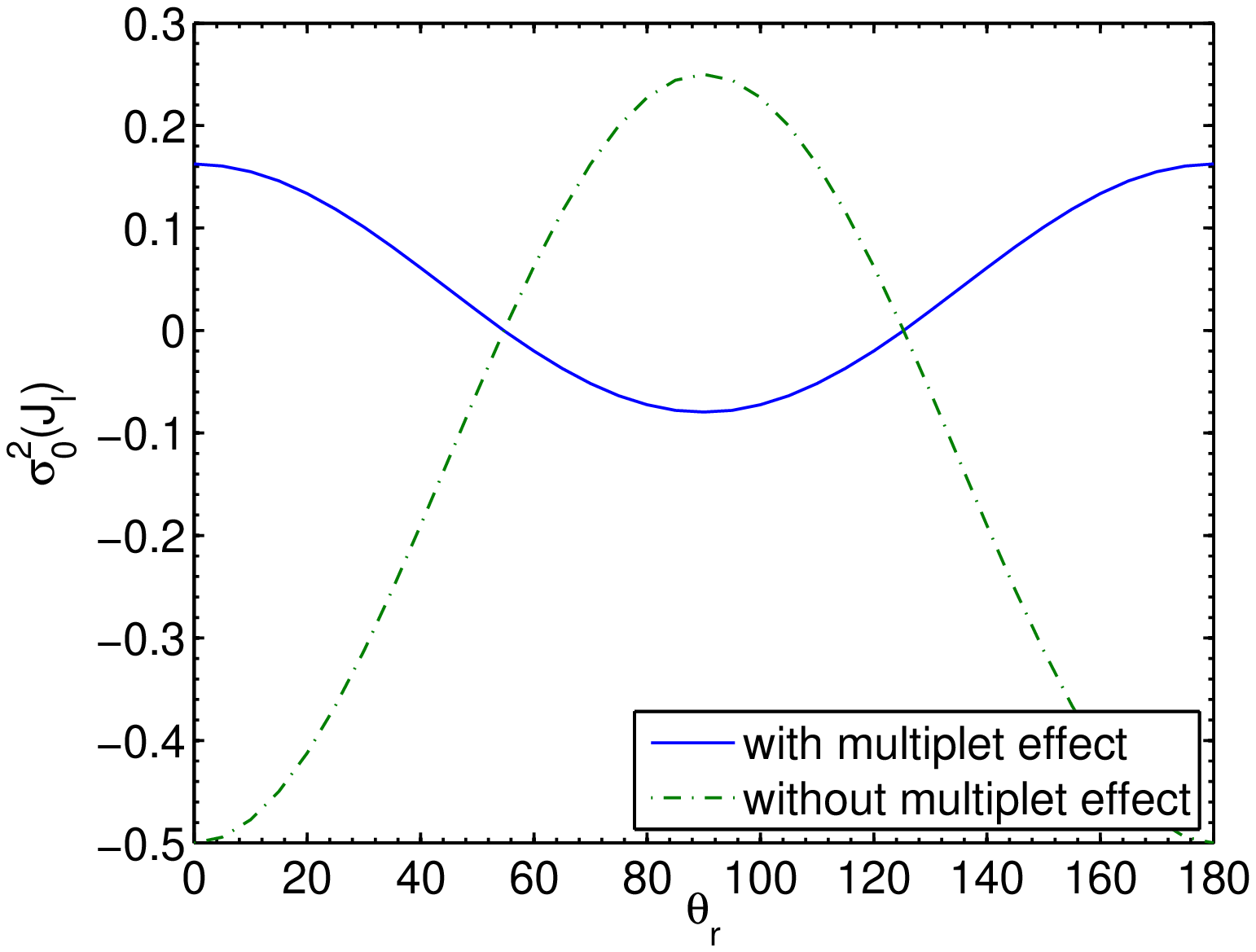}{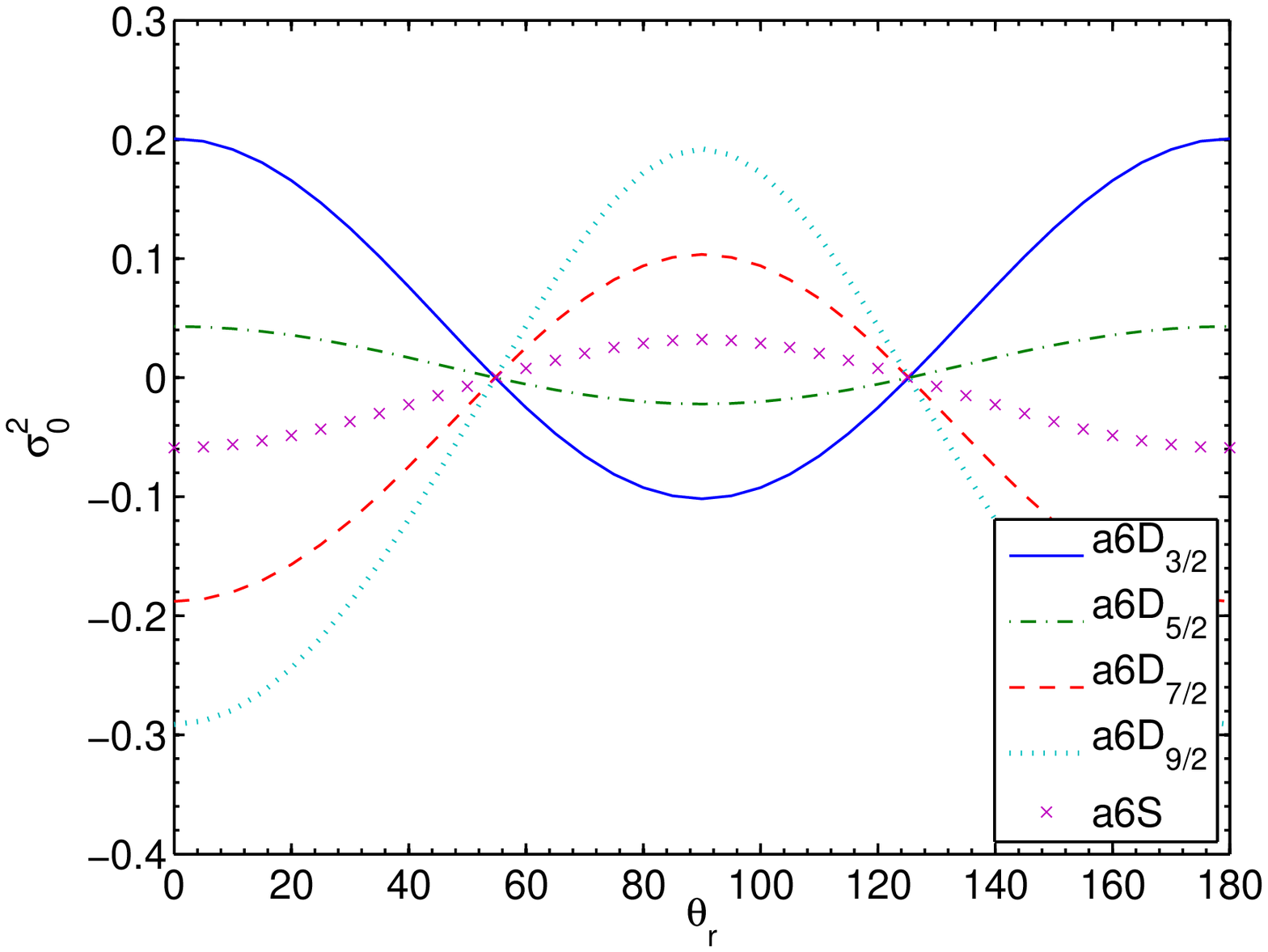}
\caption{{\it Left}: The normalized density tensor components $\sigma^2_0=\rho^{2}_0/\rho^0_0$ of the ground state of S II, dash-dot line is the result if we only consider the transition ($4S^o_{1/2}-4P_{3/2}$) itself; solid line is the actual result with the multiplet effect taken into account, the transitions between the ground state $4S^o_{3/2}$ and other upper levels $4P_{3/2, 5/2}$ reduce the degree of alignment; {\it Right}: The normalized dipole density components $\sigma^2_0=\rho^2_0/\rho^0_0$ of the ground state $a6S$ ('x' line) and metastable state $a6D_{3/2,5/2,...9/2}$ (other lines) of Cr II.}
\label{ground}
\end{figure}

The fact that $\nu_L\gg \tau_R^{-1}$ indicates  that the coefficients of the diagonal terms with $q\neq 0$ in Eq.(\ref{lowlevel}) are much larger than the other coefficients. This implies all Zeeman coherence components ($q\neq 0$) disappear. It is expected that there is no Zeeman coherence in the magnetic quantization frame as magnetic mixing is fast. Owing to the selection rule of "3j" symbols (see App.\ref{angmtmcpl}), only ${\bar J}^{0,2}_{Q=0}$ components appear and they are determined by the polar angle $\theta_r$ of the radiation (Eq.\ref{irredradia}, Fig.\ref{radiageometry}). As a result, $\rho^{2,4}_0$ are real quantities and they are independent of the azimuthal angle $\phi_r$ (see Eq.\ref{irredradia}). Physically this results from fast procession around magnetic field. Their dependence on polar angle is shown in Fig.\ref{ground}. To calculate the alignment for the multiplet, all transitions should be counted according to their probabilities even if one is interested in only one particular line. This is because they all affect the ground populations and therefore the degree of alignment and polarization. This means that in practical calculations, a summation should be taken over all the three upper levels $J_u=1/2,3/2,5/2$ in Eq.(\ref{lowlevel}). Inserting the values of $J_u,J_l$, $k,k'$, $K$ and ${\bar J}^K_Q$ (Eq.\ref{irredradia}) into Eq.(\ref{lowlevel}), we get three linear equations for $\rho^{0,2}_0(J_l)$. By solving them, we obtain  
\be
\sigma^2_0(J_l)=\frac{1.0481-3.1442 \cos^2\theta_r}{0.264 \cos^2\theta_r-13.161}
\label{S2ground}
\ee
where the floating point numbers are approximate. The same is true with other results in the paper; we take $\geq 4$ significant digits. To demonstrate the effect of multiplet, we also calculate the alignment resulting from only one transition $3/2\rightarrow 1/2$. The comparison is given in Fig.\ref{ground}. Apparently, multiple transitions reduce the degree of alignment.

\begin{figure}
\includegraphics[%
  width=0.45\textwidth,
  height=0.3\textheight]{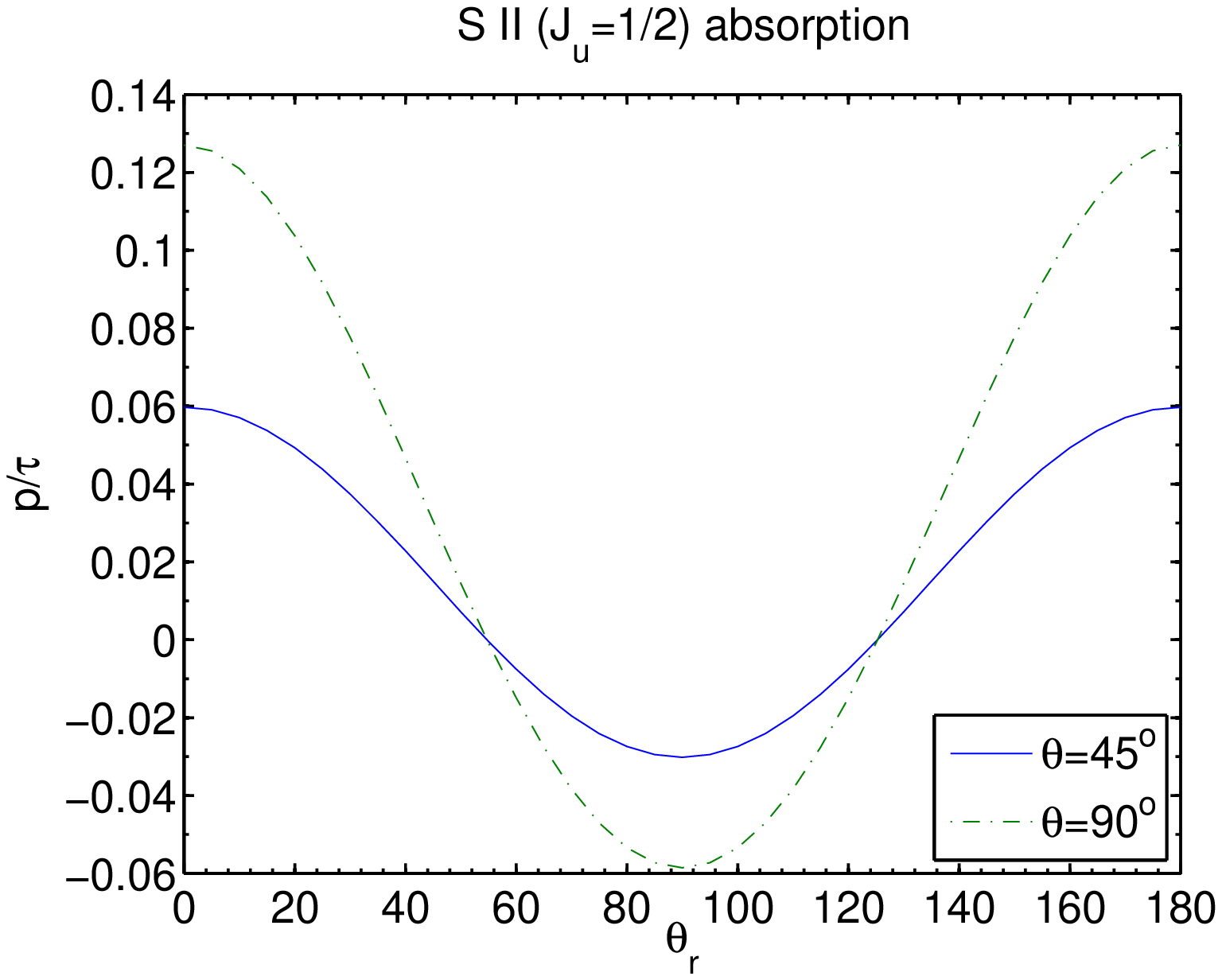}
\includegraphics[%
  width=0.45\textwidth,
  height=0.3\textheight]{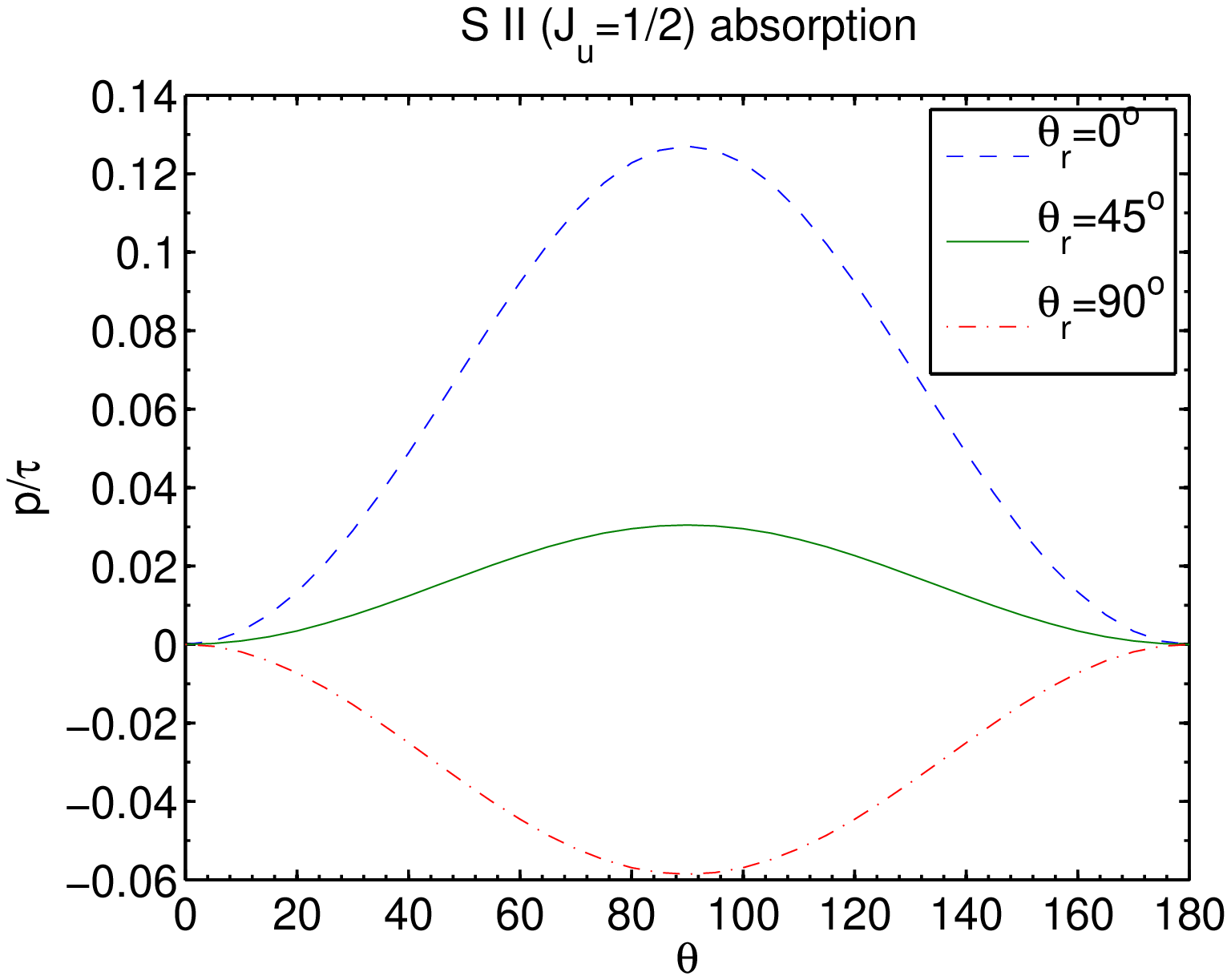}
\caption{Degree of polarization of S II absorption lines vs. $\theta_r$, the angle between magnetic field and direction of pumping source ({\it left}); $\theta$, the angle between magnetic field and line of sight ({\it right}).}
\label{polarization}
\end{figure}

Substituting Eq.(\ref{S2ground}) into Eq.(\ref{absorb}), we get the polarization of S II absorption line,
{\scriptsize
\be
\frac{P}{\tau}=\frac{( 1.5722-4.7163\cos^2\theta_r)\sin^2\theta w^2_{J_lJ_u}}{(1.4822-4.4466\cos^2\theta_r)-(1.0481-3.1442 \cos^2\theta_r)(1-1.5\sin^2\theta)w^2_{J_lJ_u}}
\ee}
Fig.\ref{polarization} shows the dependence of the degree of polarization on $\theta_r$ and $\theta$. Fig.\ref{S2contour}{\it left} is the corresponding contour plot. 
We see that the polarization in absorption is either parallel or perpendicular to the plane of sky. The polarization changes sign at the Van Vleck angle $54.7^o$ (Van Vleck 1925, House 1974). This happens because of coupling of the two oscillators in the $x,y$ direction. These two oscillators share energy due to Larmor precession around the magnetic field (in the direction of $z"$ axis). Also, as can be seen from the dipole component of  radiation tensor ${\bar J}^2_0$ of unpolarized radiation, ${\bar J}^2_0\propto{\cal J}^2_0\propto (1-1.5\sin^2\theta_r)/2$. Therefore the dipole component of the density matrix $\rho^2_0$ , which is proportional to ${\bar J}^2_0$ (Eq.\ref{lowlevel}, \ref{S2ground}), changes from parallel to perpendicular to the magnetic field at Van Vleck angle (Fig.\ref{ground}). As the result the polarization of the absorption line also changes according to Eq.(\ref{absorb}). {\it This turnoff at the Van Vleck angle is a general feature regardless of the specific atomic species as long as the background source is unpolarized and it is in the atomic realignment regime.} 

The absorption shown in Fig.\ref{polarization},\ref{S2contour}{\it left} exhibits variations with $\theta_r$ and $\theta$. The maximal polarization (around 12 percent) is expected for the
S II ($J_u=1/2$) component. At $\theta=90^o$, the observed polarization reaches a maximum for the same $\theta_r$ and alignment, which is also expected from Eq.(\ref{absorb}). This shows that atoms are indeed realigned with respect to magnetic field so that the intensity difference is maximized parallel and perpendicular to the field (Fig.\ref{nzplane}{\it right}). At $\theta=0^o, 180^o$, the absorption polarization is zero according to Eq.(\ref{absorb}). Physically this is because the precession around the magnetic field makes no difference in the $x,y$ direction when the magnetic field is along the line of sight (Fig.\ref{nzplane}{\it right}). 

\begin{figure}
\includegraphics[%
  width=0.45\textwidth,
  height=0.3\textheight]{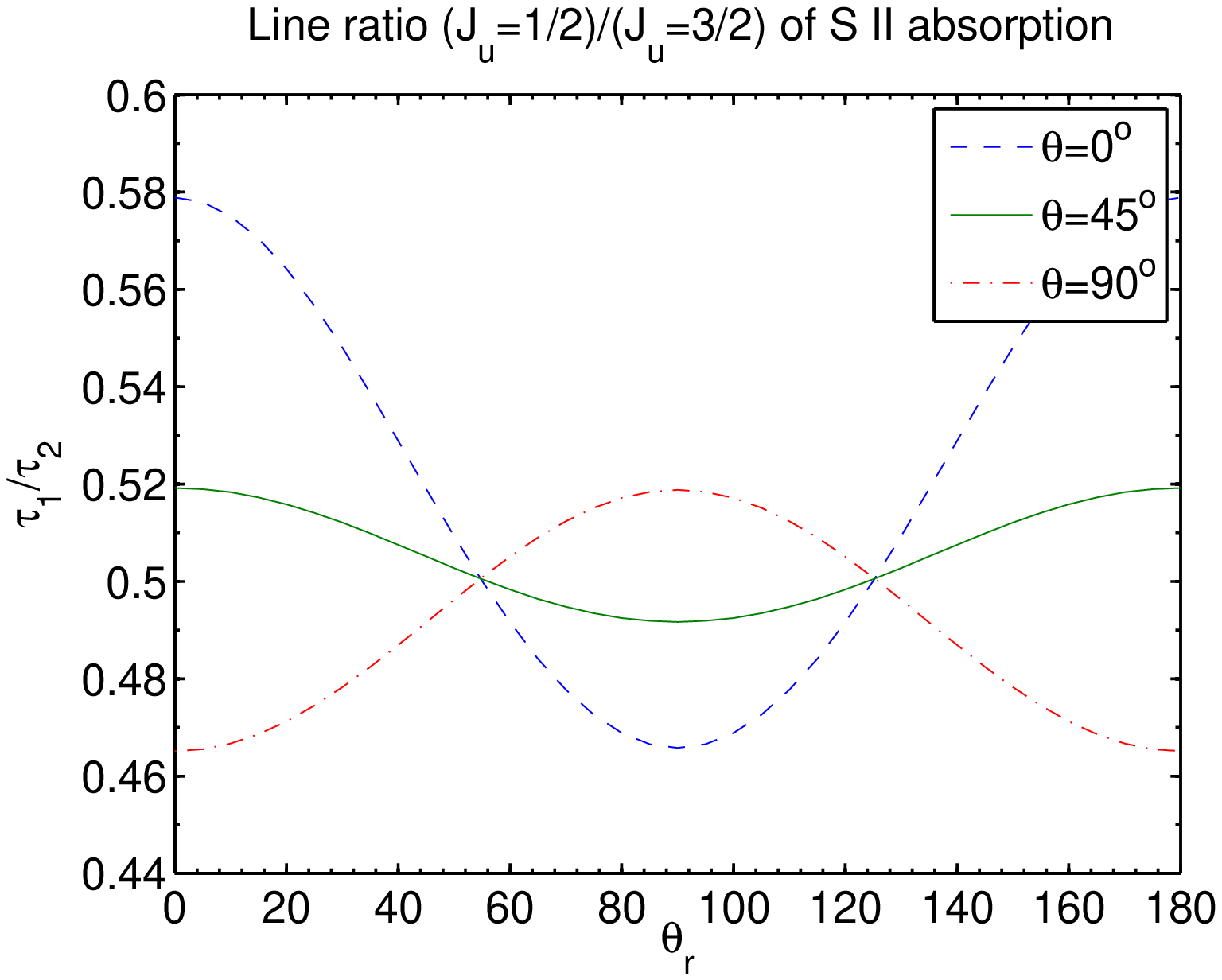}
\includegraphics[%
  width=0.45\textwidth,
  height=0.3\textheight]{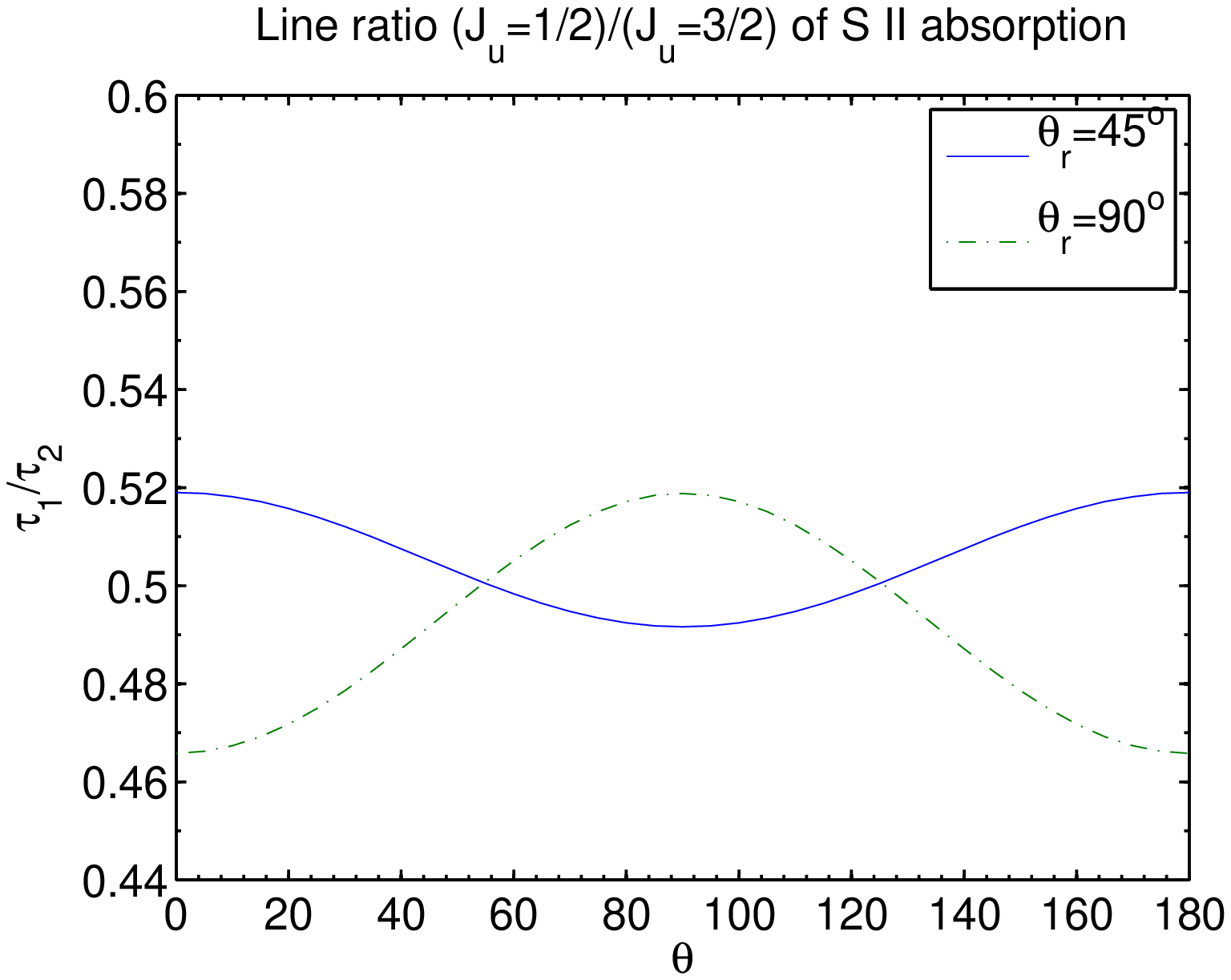}
\caption{Optical depth ratios of S II absorption lines vs. $\theta_r$, the angle between magnetic field and direction of pumping source ({\it left}); $\theta$, the angle between magnetic field and line of sight ({\it right}).}
\label{lineratio}
\end{figure}

\begin{figure}
\includegraphics[%
  width=0.45\textwidth,
  height=0.3\textheight]{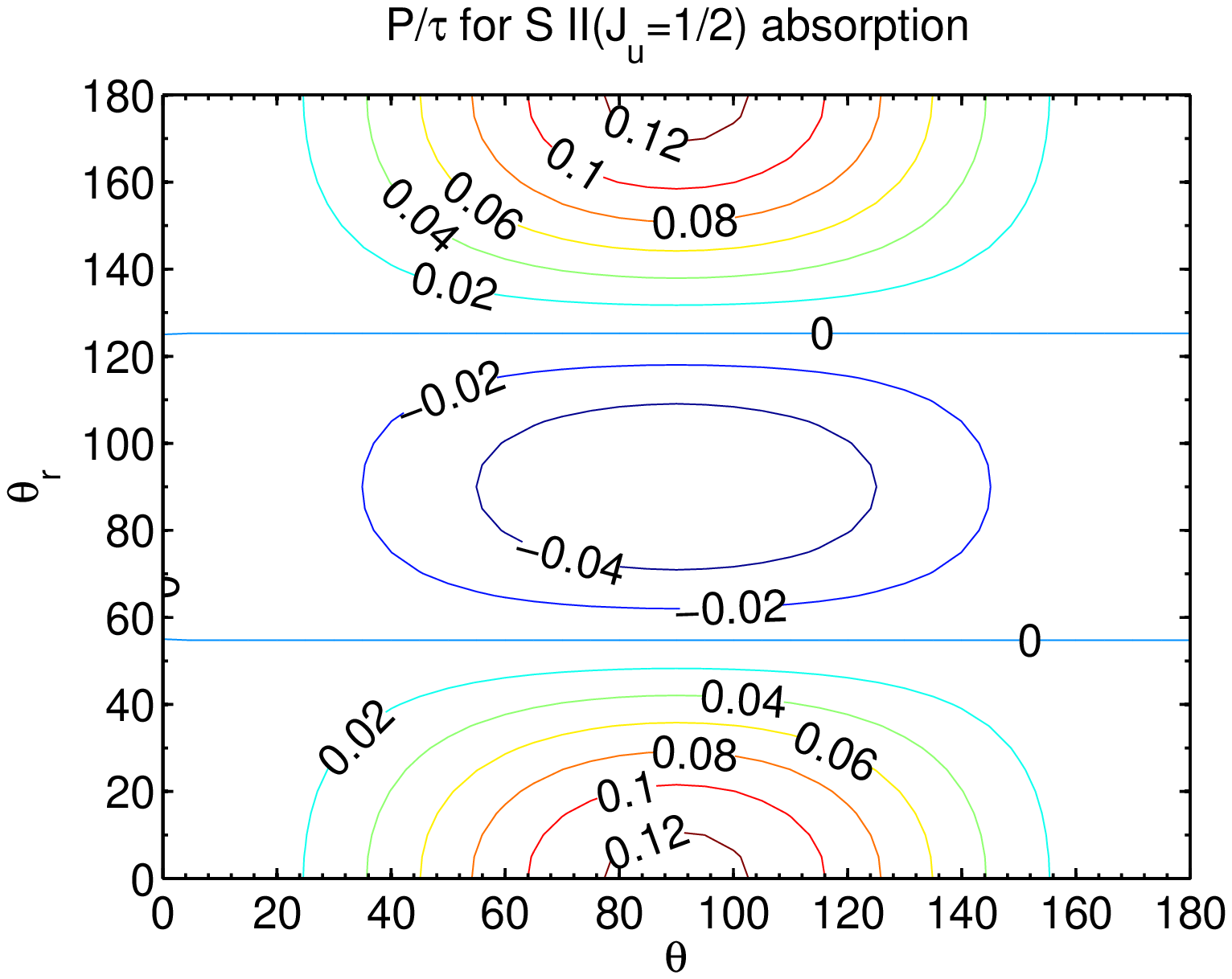}\includegraphics[%
  width=0.45\textwidth,
  height=0.3\textheight]{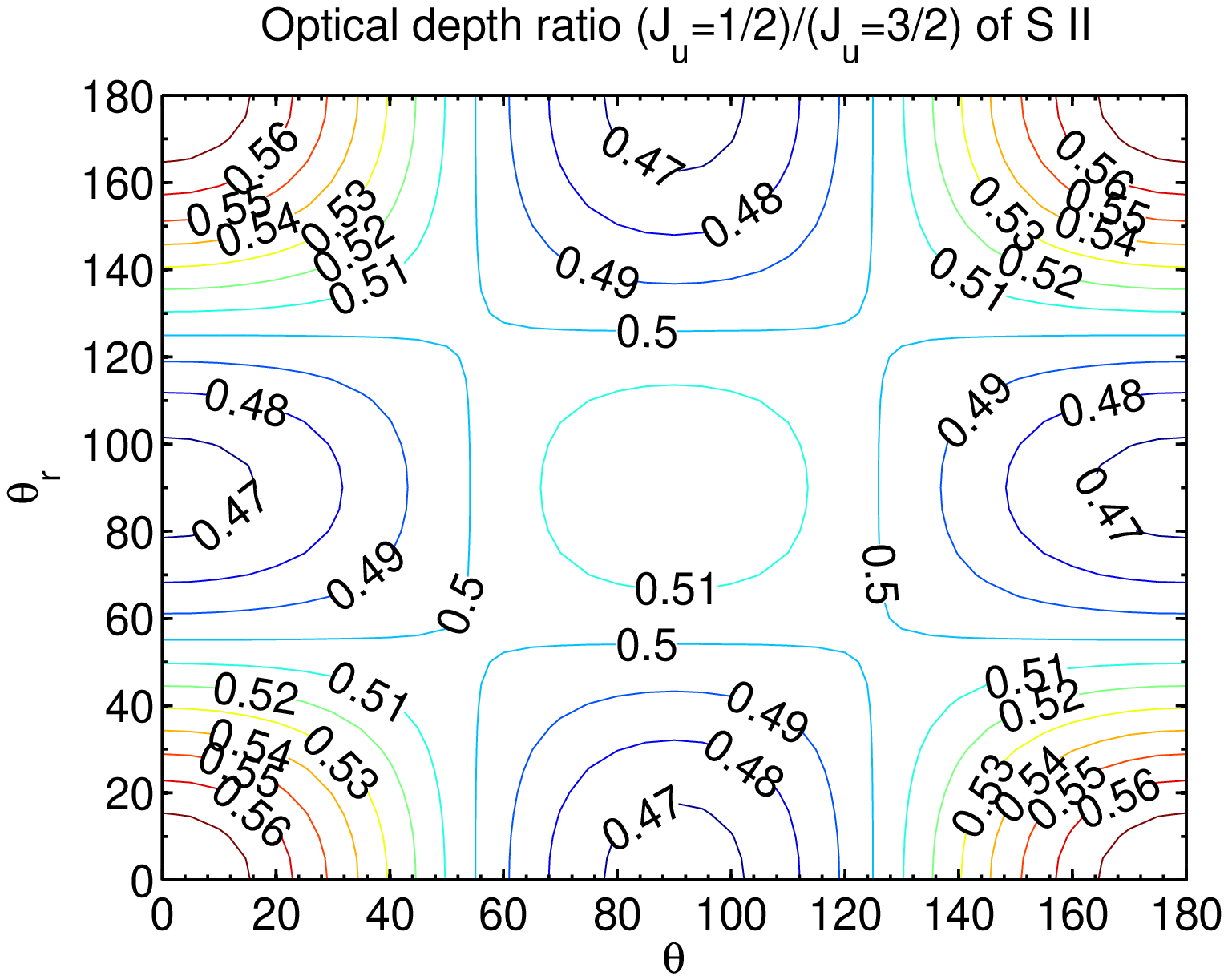}
\caption{The contour graphs of S II polarization and line ratio. They are determined by the dipole component of density matrix $\sigma^2_0(\theta_r)$ and the direction of observation $\theta$ (Eqs.\ref{absorb},\ref{tauratio}). Since the $\theta_r$ (Fig.\ref{geometry}) dependence is solely from $\sigma^2_0$, they exhibit similar pattern in the vertical direction.}
\label{S2contour}
\end{figure}

The optical depth also varies as a result of the alignment (see Eq.\ref{tauratio}). We plot the optical depth ratios of line ($J_u=1/2$) to line ($J_u=3/2$) (see Fig.\ref{lineratio} and right panel of Fig.\ref{S2contour}). For optically thin lines, the equivalent width ratio is equal to that of optical depths. Note that when the magnetic field is parallel to the incident radiation, the magnetic field does not have an impact. The corresponding line ratios without magnetic realignment are equal to those values at $\theta_r=0^o$ in the graphs.

The fact that absorption polarization can be only parallel or perpendicular to the magnetic field in the plane of sky has an important implication. It means that the direction of polarization traces the projection of magnetic field in the plane of sky (the angle $\phi_B$, see Fig.\ref{radiageometry}) within a $90^o$ uncertainty. This is a very {\it ``easy"} information. Furthermore if we get two measurables, the degree of polarization and the ratio of a doublet\footnote{S II is a triplet, for which we can get three degrees of polarizations and two line ratios. However, the degrees of polarizations are proportional to each other as they have the same dependence on $\sigma^2_0(J_l)$ (see Eq.\ref{absorb}). So do the line ratios. Thus only by combining one degree of polarization and one line ratio, we can obtain the two angles.}, we shall be able to extrapolate both $\theta_r$ and $\theta=\theta_B$ (see Fig.\ref{radiageometry}). With $\theta_r$ known, we can remove the $90^o$ degeneracy of $\phi_B$ and thus obtain the 3D information of the magnetic field ($\theta_B,\phi_B$).

We consider a general case where the pumping source does not coincide with the object whose absorption is measured. 
If the radiation that we measure is also the radiation that aligns the atoms, the direction of pumping source coincides with line of sight, {\it i.e.}, $\theta=\theta_r$ (Fig.\ref{radiageometry}).

{\bf Cr I} has a ground state $7S_3$ and excited state $7S_{2,3,4}$. The computing procedure is similar to that for S II. The difference is that the density matrix for the Cr I ground state has seven components with $k=3-3, 3-2,...,3+3=0,1,2,3,4,5,6,7$. We only need to consider even components $\rho^{0,2,4,6}_0$. Following the same procedure as described for S II, we get the following results:  
\bea
\left[\begin{array}{c}\sigma^2_0(J_l)\\\sigma^4_0(J_l)\\\sigma^6_0(J_l)\end{array}\right]&=&\left(
\begin{array}{l}
 35.829\cos 2\theta_r-12.593\cos 4\theta_r+0.253\cos6\theta_r+1.932\\
 0.671\cos 2\theta_r+1.2648\cos 4\theta_r-0.6092\cos 6\theta_r+1.7263\\
 -1.0902\cos 2\theta_r-0.5032\cos 4\theta_r-0.2516\cos6\theta_r-0.5404
\end{array}
\right)/(-322\cos 2\theta_r+1.24\cos 4\theta_r+568).
\eea
By inserting these results into Eq.(\ref{absorb}) or (\ref{generalabsorb}), one can get the polarizations of Cr I absorption lines.

\subsection{More complex structure with multiple lower levels: C II, Si II, O I, S I,  \& Cr II}
\subsubsection{Strong Pumping}
\label{strongp}
In this section we discuss atoms with more complex structure. As illustrated in Table 1, C II has not only multiple upper levels, but also two levels $2P^o_{1/2, 3/2}$ on the ground state. The level $2P^o_{1/2}$ has altogether two magnetic projections $M_J=\pm 1/2$, and is thus not alignable and has only one density tensor $\rho^0_0$. Level $2P^o_{3/2}$ has density components $\rho^{0,2}_0$. We shall consider the five lines $1036.34\AA (2P^o_{1/2}\rightarrow 2S_{1/2}), 1037.02\AA (2P^o_{3/2}\rightarrow 2S_{1/2}), 1334.53\AA (2P^o_{1/2}\rightarrow 2D_{3/2}), 1335.66\AA (2P^o_{3/2}\rightarrow 2D_{3/2}), 1335.71 (2P^o_{3/2}\rightarrow 2D_{5/2})$, which are observed in QSOs. All the transitions between upper levels and the ground state should be counted. There is also a magnetic dipole transition between the two ground levels. Although its transition probability is very low, it can be comparable to the optical pumping rate in regions far from any radiation source. Depending on how far away the radiation source is, there can be two regimes divided by the boundary where the magnetic  dipole radiation rate $A_m$ is equal to the pumping rate $\tau_R^{-1}$. Inside the boundary, the optical pumping rate is much larger than the M1 transition rate $A_m$ so that we can ignore the magnetic dipole radiation as a first order approximation. Further out, the magnetic dipole transition is faster than optical pumping so that it can be assumed that most atoms reside in the lowest energy level of the ground state. For the case of C II that we consider now, the lowest level is not alignable and so we shall concentrate on the first regime, and neglect the magnetic dipole transition.

\begin{figure}
\includegraphics[%
  width=0.35\textwidth,
  height=0.28\textheight]{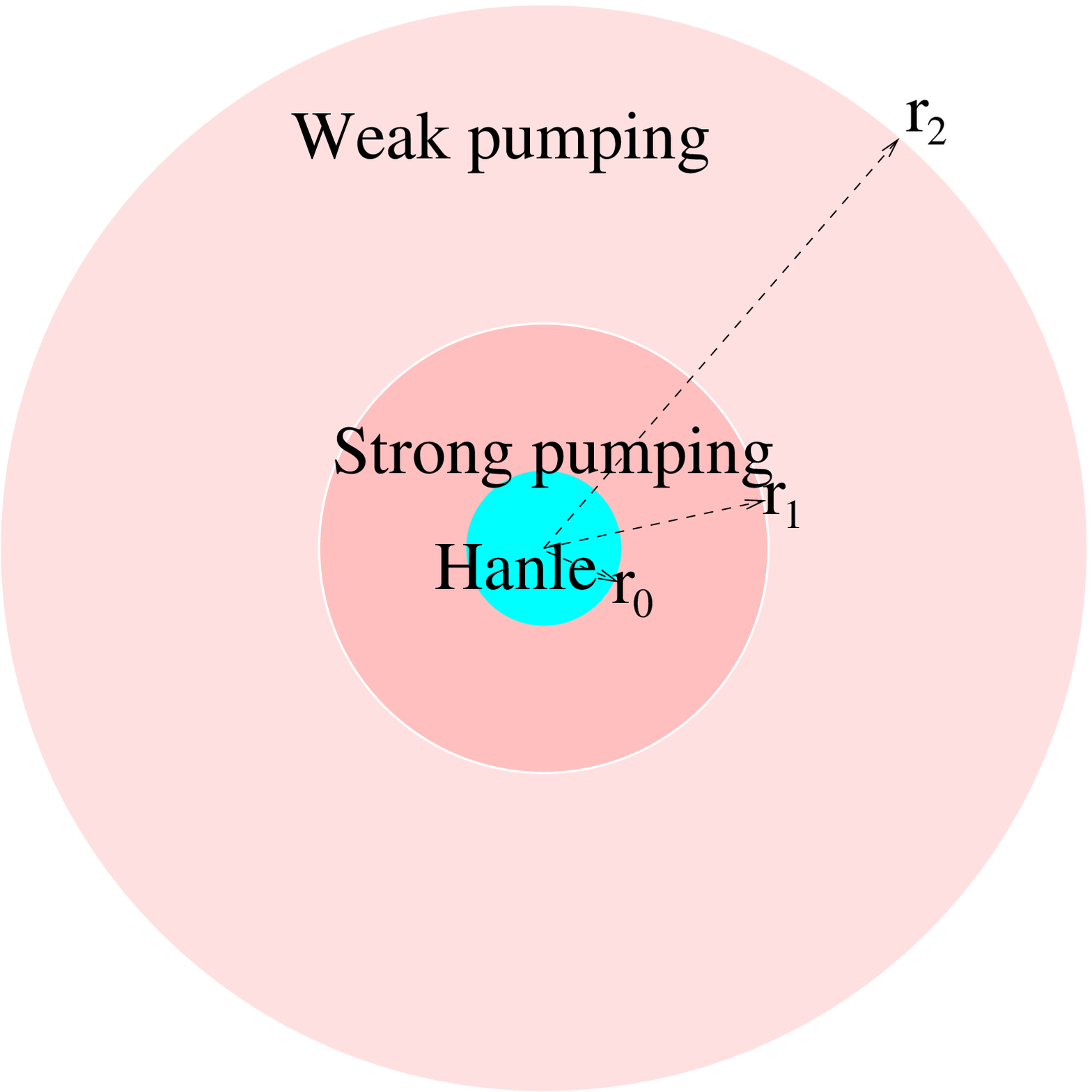}
\includegraphics[%
  width=0.64\textwidth,
  height=0.35\textheight]{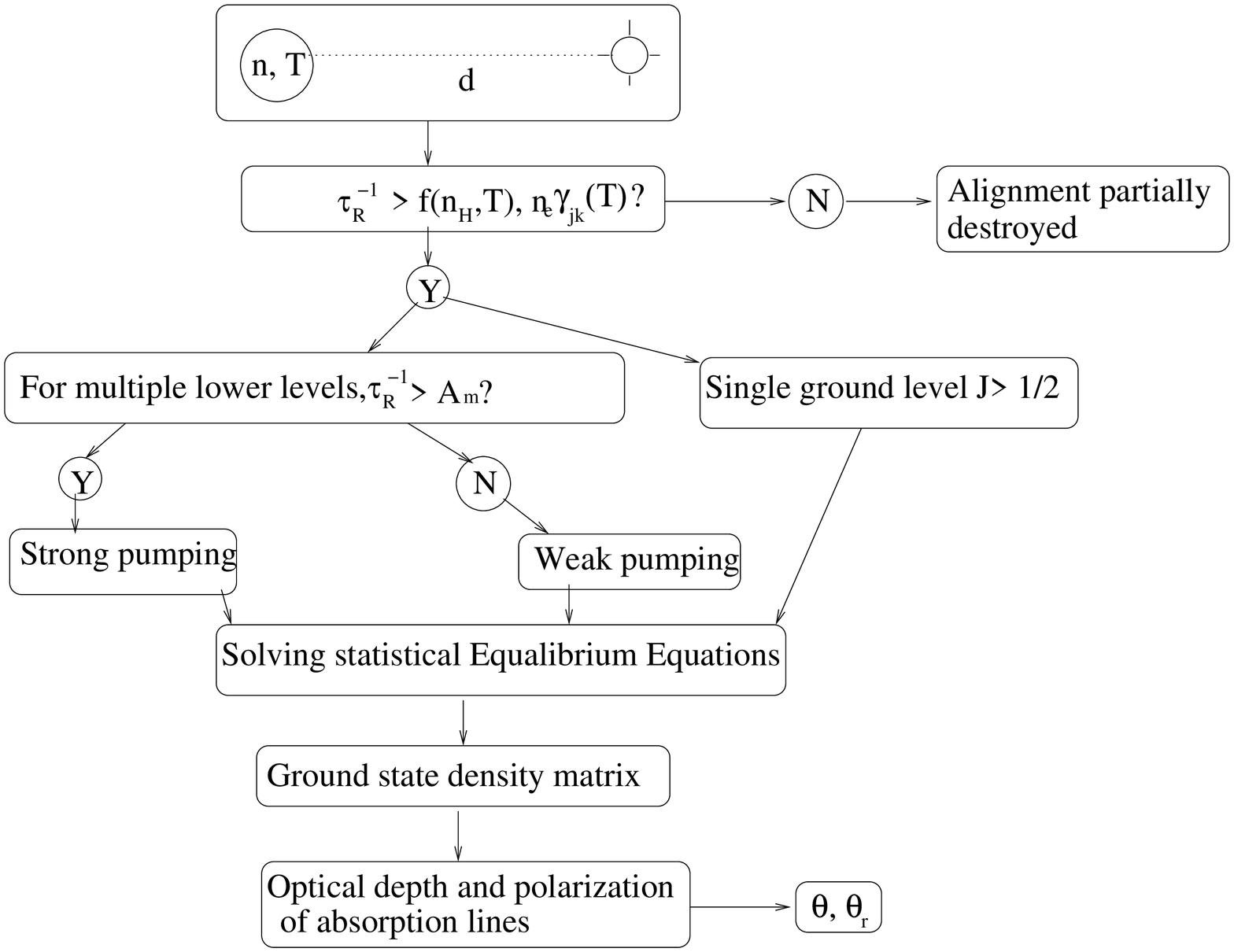}
\caption{{\it left}: a carton illustrating how the atomic pumping changes with distance around a radiation source. For circumsteller region, magnetic field is strong, such that the Hanle effect, which requires $\nu_L\sim A$, dominates. Atomic alignment applies to the much more distant ($\lesssim r_2$, see Eq.\ref{r0}) interstellar medium. Inside $r_2$, it can be further divided into two regimes: strong pumping and weak pumping, demarcated at $r_1$ (see Eq.\ref{demarcator}); {\it Right}: whether and how atoms are aligned depends on their intrinsic properties (transitional probabilities and structures) and the physical conditions: density n, temperature T and the averaged radiation intensity from the source $I_*$. If the pumping rate $\tau_R^{-1}$ is less than collisional rates, alignment is partially destroyed. Then for atoms with multiple lower levels, depending on the comparison between the pumping rate and the magnetic dipole radiation rate among the lower levels, the atoms are aligned differently. In the strong pumping case, all the alignable lower levels are aligned; on the contrary, only the ground level can be aligned in the case of weak pumping (see discussions in \S\ref{strongp}).}
\label{diagram}
\end{figure}

The two regimes are demarcated at $r_1$ (see Fig.\ref{diagram}{\it left}), where $r_1$ is defined by 
\be 
\tau_R^{-1}=B_{J_lJ_u}I_{BB}(R_*/r_1)^2=A_m,
\label{demarcator}
\ee 
and $I_{BB}, R_*$ are the intensity and radius of the pumping source. Outside the boundary where $\tau_R^{-1}\ll A_m$, C II and Si II are not alignable and so their absorption lines are unpolarized. For different radiation sources, the distance to the boundary differs. For an O type star, the distance would be $\sim$0.1pc, while for a shell star ($T_{eff}=15,000$K), it is as close as to $\sim 0.003-0.01$ pc. The absence of alignment outside the radius $r_1$
ensures that observations constrain the magnetic field topology within this radius.

 Since there are multiple transitions with frequencies that span over a wide range, the optical pumping and its resulting atomic alignment depend on the shape of the spectrum of the radiation source. Neglecting the difference within an LS multiplet, we characterize the intensity of the multiplet ($2P^o\rightarrow 2S_{1/2}$) by $I_S$, multiplet ($2P^o\rightarrow 2P_{1/2,3/2}$) by $I_P$, and that of multiplet ($2P^o\rightarrow 2D_{1/2,3/2,5/2}$) by $I_D$. In the case that other multiplets of the same terms have comparable intensities, $I_S$, $I_P$, $I_D$ are weighted summations of all the intensities of the same terms (e.g, see Eq.\ref{Si2}),
\bea
I_S&=&I_{1036.3}\nonumber\\
I_P&=&I_{904.1}\nonumber\\
I_D&=&I_{1334.5}
\label{C2}
\eea 
These are intensities per unit frequency. So are other intensities in the paper. We use wavelengths (in unit of $\AA$) rather than frequencies in the subscripts as the lines are in optical and UV band. Note that, due to absorption from the Lyman series, those lines with wavelengths shorter than $912\AA$ can be ignored in the case of distant radiation pumping sources. We obtain for {\bf C II}:
{\scriptsize
\bea
\rho^{0}_0(J_l=3/2)&=&\varrho^0_0\left[\left(0.44 I_D^2+0.39
   I_P I_D-1.65 I_S
   I_D-7.06 I_P^2-7.37
   I_P I_S\right) \cos
   ^4\theta_r+\left(-6.54
   I_D^2-47.81 I_P
   I_D-17.56 I_S
   I_D+25.88I_P^2\right.\right.\nonumber\\
&+&\left.\left.12.79   I_P I_S\right) \cos ^2\theta_r+31.33
   I_D^2+408.70 I_P^2+131.63
   I_S^2+302.23 I_D
   I_P+137.29 I_D
   I_S+505.48 I_P I_S\right] 
\nonumber\\
\rho^{2}_0(J_l=3/2)&=& \varrho^0_0\left[-0.72 I_D^2-3.77 I_PI_D+8.88 I_S
   I_D+12.75 I_P^2+32.91
   I_S^2+\left(2.15
   I_D^2+11.32I_P
   I_D-26.65 I_S
   I_D-38.25 I_P^2-98.72
   I_S^2\right.\right.\nonumber\\
&-&\left.\left.275.40 I_P I_S\right) \cos^2\theta_r+91.80 I_P I_S\right],
\eea}
where 
{\scriptsize 
\bea
\varrho^0_0&=&\rho^0_0(J_l=1/2)(\left(-0.60 I_D^2+4.92 I_P   I_D+16.20I_S I_D-9.98I_P^2-52.29 I_S^2-60.48 I_P   I_S\right)\cos ^4\theta_r+\left(-4.0   I_D^2-21.13 I_P   I_D\right.\nonumber\\
&-&\left.23.98 I_S   I_D+14.97 I_P^2+34.86   I_S^2+45.88 I_P I_S\right) \cos   ^2\theta_r+21.95 I_D^2+159.68   I_P^2+87.15 I_S^2+135.08 I_D   I_P+98.79 I_D   I_S+239.82 I_P I_S).\nonumber
\eea}

{\bf Si II} has the same structure as C II does.  Though there are many lines in the range from $989.87\AA$ to $1533.43\AA$, their upper levels belong to only three terms: $2S_{1/2}, 2P_{1/2,3/2}$ and $2D_{1/2,3/2,5/2}$, the
same as C II. The difference of its alignment from C II is solely due to the difference in the line strengths of the multilevels involved, 
{\scriptsize
\bea
\rho^{0}_0(J_l=3/2)&=&\varrho^0_0\left[
 \left(-0.68 I_D^2-0.16
   I_P I_D+0.46 I_S
   I_D+0.74 I_P^2+0.54
   I_P I_S\right) \cos
   ^4\theta_r+\left(10.09
   I_D^2+19.32I_P
   I_D+4.92 I_S
   I_D-2.72 I_P^2\right.\right.\nonumber\\
&-&\left.\left.0.93 I_P I_S\right) \cos ^2\theta_r-48.31
   I_D^2-42.97 I_P^2-6.65
   I_S^2-121.92 I_D
   I_P-38.36 I_D
   I_S-36.82 I_P I_S\right] \nonumber\\
\rho^{2}_0(J_l=3/2)&=&\varrho^0_0\left[\left(I_D^2+2.0193I_P
   I_D+0.2196 I_S
   I_D-0.1412 I_P^2-0.1750
   I_S^2-0.7037 I_P I_S\right)
   \left(9.497-28.490 \cos ^2\theta_r\right)\right],
\eea}
where 
{\scriptsize 
\bea
\varrho^0_0&=&\rho^0_0(J_l=1/2)/\left[
\left(11.59 I_D^2-11.96 I_P
   I_D-11.73 I_S I_D+1.05
   I_P^2+2.63 I_S^2+4.40 I_P
   I_S\right) \cos ^4\theta_r+\left(-0.920
   I_D^2+15.182 I_P I_D\right.\right.\nonumber\\
&+&\left.\left.11.506 I_S
   I_D-1.573 I_P^2-1.756
   I_S^2-3.338 I_P I_S\right) \cos
   ^2\theta_r-32.783 I_D^2-16.781
   I_P^2-4.389 I_S^2-55.5989
   I_D I_P-28.387 I_D I_S-17.447
   I_P I_S\right],\nonumber 
\eea}
and
\bea
I_S&=&(I_{1533.4} S_{1533.4}+I_{1309.3} S_{1309.3}+I_{1023.7}S_{1023.7})/S_{1533.4},\nonumber\\
I_P&=&I_{1194.5},\nonumber\\
I_D&=&(I_{1816.9} S_{1816.9}+I_{1264.7} S_{1264.7}+...)/S_{1816.9},
\label{Si2}
\eea 
in which $S_{\lambda}\propto gB$ is the line strength of the line with wavelength $\lambda$. The typical values of these intensities $I_S, I_P, I_D$ are listed in Table~\ref{intensity}. If we use these density tensors with Eqs(\ref{absorb},\ref{tauratio}), we get the absorption coefficients and the corresponding polarizations and line ratios. 
{\large
\begin{table}
\begin{tabular}{|c|c|c|c|c|c|c|c|}
\hline
\hline
&\multicolumn{2}{|r}C II~~~~& Si II&\multicolumn{1}{|c}{O I}&S I&\multicolumn{2}{r|}{Ti II~~~}\\
\hline
ISRF&$10^{14}I_S$&0.047718&0.1574&0.10051&0.10697&$10^{14}I_F$&34.445\\
\cline{3-6}\cline{8-8}
&$10^{14}I_P$&0&0.077493&0&0.091174&$10^{14}I_G$&39.734\\
\cline{3-6}\cline{8-8}
&$10^{14}I_D$&0.096898&0.10106&0.062759&0.14683&$10^{14}I_D$&28.295\\
\hline
shell star&$I_S/W $&0.3429&2.8065&1.1893&3.3543&$I_F/W$&6.3717\\
\cline{3-6}\cline{8-8}
&$I_P/W$&0&0.7597&0&1.1146&$I_G/W$&6.39\\
\cline{3-6}\cline{8-8}
&$I_D/W$&1.2704&1.0292&0.43198&2.4212&$I_D/W$&6.3021\\
\hline
O-type star&$I_S/W$&237.11&342.79&255.94&172.05&$I_F/W$&81.5464\\
\cline{3-6}\cline{8-8}
&$I_P/W$&0&230.45&0&222.49&$I_G/W$&76.4647\\
\cline{3-6}\cline{8-8}
&$I_D/W$&218.89&251.90&390.29&314.73&$I_D/W$&88.5257\\
\hline
\hline
\end{tabular}
\caption{The line intensities ($10^{-4}{\rm ergs}/{\rm cm^2 \cdot s \cdot Hz \cdot sr}$) for pumpings of C II, Si II, O I, S I and Ti II (see Eq.\ref{C2},\ref{Si2},\ref{O1},\ref{S1}) in typical astrophysical environment: interstellar radiation field (ISRF), shell star ($T_{eff}=15,000$K), O-type star ($T_{eff}=50000$K). W is the dilution factor $\left(R_*/r\right)^2$.}
\label{intensity}
\end{table}
}
O I and S I have ground states $3P_{0,1,2}$ and upper levels $3S^o_1, 3D^o_{1,2,3}$. The ground level is $3P_2$. Thus, unlike C II and Si II, O I and S I are alignable in both regimes regardless of $\tau_R^{-1}> {\rm or}<A_m$. The degree of alignment, however, differs in the two regimes. The case for weak pumping will be presented in \S\ref{weakp}. In the case of strong pumping, that we consider here, both levels of $3P_{1,2}$ are alignable. Eq.(\ref{lowlevel}) is then a five dimensional matrix with $\rho^{0,2}(J_l=1)$ and $\rho^{0,2,4}(J_l=2)$. The explicit generic solution is very lengthy. Here as an example, we give the results for pumping by a shell star (a black body source of $T=15000$K, see Table~\ref{intensity}).
For {\bf O I}:
{\scriptsize
\bea
\begin{array}{l}
\rho^{0}_0(J_l=1)\\
\rho^{2}_0(J_l=1)\\
\rho^{0}_0(J_l=2)\\
\rho^{2}_0(J_l=2)\end{array}&=&\varrho^0_0
\left(
\begin{array}{l}
 0.0851\cos ^{14}\theta_r-0.4823 \cos
   ^{12}\theta_r-1.3183 \cos ^{10}\theta_r+7.2195
   \cos ^8\theta_r+4.7687 \cos ^6\theta_r-20.6145\cos
   ^4\theta_r-12.0874\cos ^2\theta_r+6.9951\\
 -1.1006\cos ^{12}\theta_r+3.8022\cos
   ^{10}\theta_r+3.4199 \cos ^8\theta_r-12.6453
   \cos ^6\theta_r-4.4752\cos ^4\theta_r+7.0292
   \cos ^2\theta_r-1.4338 \\
 0.1051\cos ^{14}\theta_r-0.1746 \cos
   ^{12}\theta_r-2.8148\cos ^{10}\theta_r+8.0475\cos ^8\theta_r+7.3564 \cos ^6\theta_r-26.7061   \cos ^4\theta_r-15.9694 \cos ^2\theta_r+9.1487 \\
 1.3136 \cos ^{12}\theta_r-5.3551 \cos
   ^{10}\theta_r-2.2558\cos ^8\theta_r+18.1327 \cos
   ^6\theta_r+4.1870 \cos ^4\theta_r-8.9711\cos
   ^2\theta_r+1.9016
\end{array}
\right)
\eea},
where {\scriptsize $\varrho^0_0=\rho^0_0(J_l=0)/0.3989 \cos
   ^{14}\theta_r-1.6228 \cos
   ^{12}\theta_r-1.3920 \cos
   ^{10}\theta_r+8.5950\cos
   ^8\theta_r+2.7734 \cos
   ^6\theta_r-14.5724 \cos
   ^4\theta_r-5.7430 \cos   ^2\theta_r+3.8703)$.}
For {\bf S I}: 
{\scriptsize
\bea
\begin{array}{l}
\rho^{0}_0(J_l=1)\\
\rho^{2}_0(J_l=1)\\
\rho^{0}_0(J_l=2)\\
\rho^{2}_0(J_l=2)\end{array}=\varrho^0_0
\left(
\begin{array}{l}
 \cos ^{14}\theta_r-2 \cos
   ^{12}\theta_r-39 \cos ^{10}\theta_r-8
   \cos ^8\theta_r+596 \cos ^6\theta_r+1437
   \cos ^4\theta_r-3303 \cos ^2\theta_r-10797 \\
 -4.9\cos ^{12}\theta_r+4.3\cos
   ^{10}\theta_r+108.4 \cos ^8\theta_r+213.6 \cos
   ^6\theta_r-814.1\cos ^4\theta_r-2929.5 \cos
   ^2\theta_r+1057.7 \\
 \cos ^{14}\theta_r-71 \cos ^{10}\theta_r-39
   \cos ^8\theta_r+851 \cos ^6\theta_r+2680
   \cos ^4\theta_r-4892 \cos ^2\theta_r-14133 \\
 7.02\cos ^{12}\theta_r-13.92 \cos
   ^{10}\theta_r-112.56\cos ^8\theta_r-234.34
   \cos ^6\theta_r+1169.11 \cos ^4\theta_r+2002.21
   \cos ^2\theta_r-787.19
\end{array}
\right),
\eea}
where {\scriptsize $\varrho_0^0=\rho^0_0(J_l=0)/(
2.3\cos
   ^{14}\theta_r-4.4 \cos
   ^{12}\theta_r-57.5 \cos
   ^{10}\theta_r-64.3 \cos
   ^8\theta_r+657.9 \cos
   ^6\theta_r+1620.9 \cos
   ^4\theta_r-2519.7 \cos
   ^2\theta_r-6083.6)$.}
In fact, {\bf C I}, {\bf Si I} and {\bf S III} have the same ground state term $3P_{0,1,2}$, but with an opposite energy sequence, namely, $3P_0$ is the ground level. Thus they are only alignable in the strong pumping case. In this case, their alignment is similar to that of O I and S I. The results are given in App.~\ref{CSIS}.

\begin{figure}
\includegraphics[%
  width=0.45\textwidth,
  height=0.3\textheight]{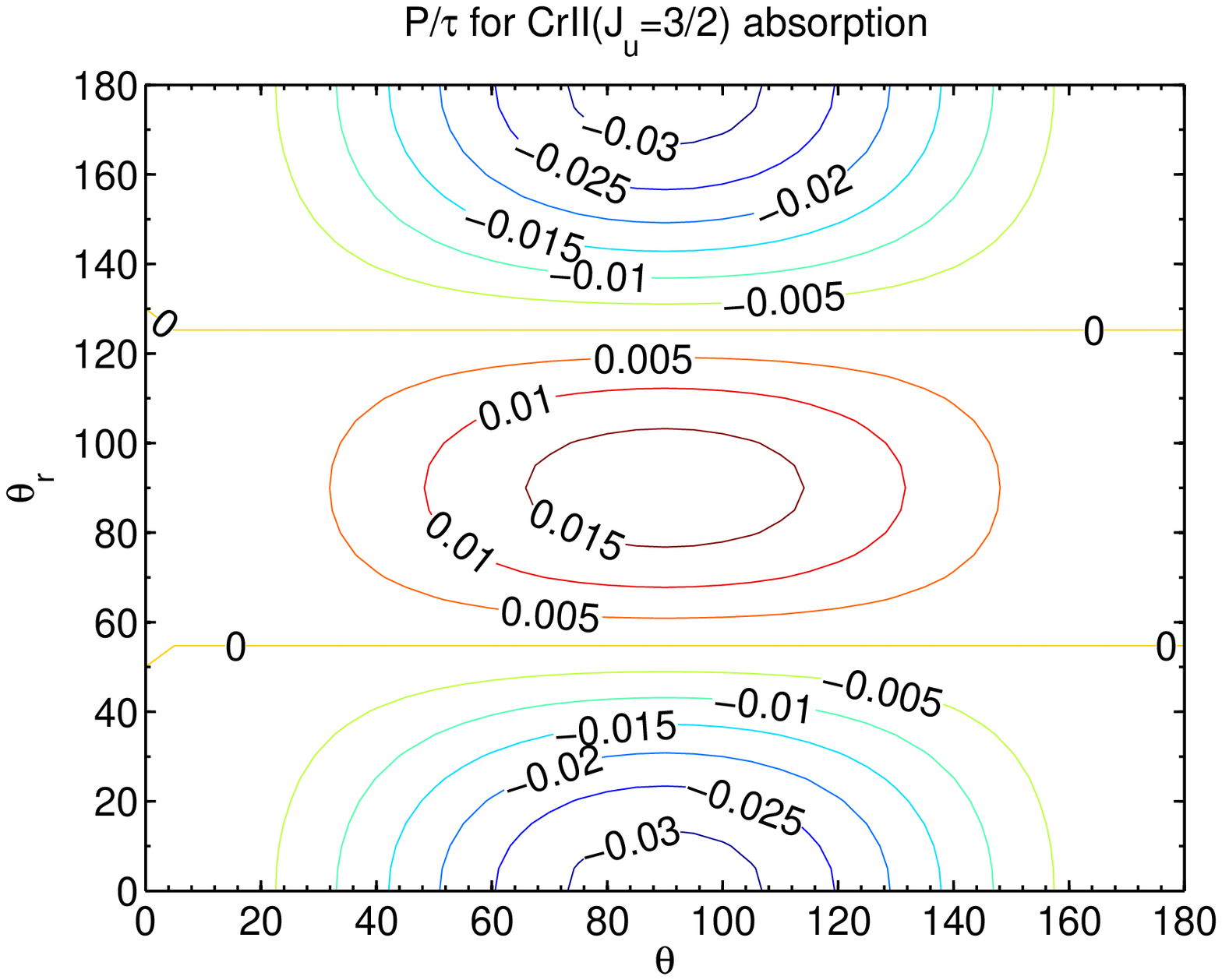}
\includegraphics[%
  width=0.45\textwidth,
  height=0.3\textheight]{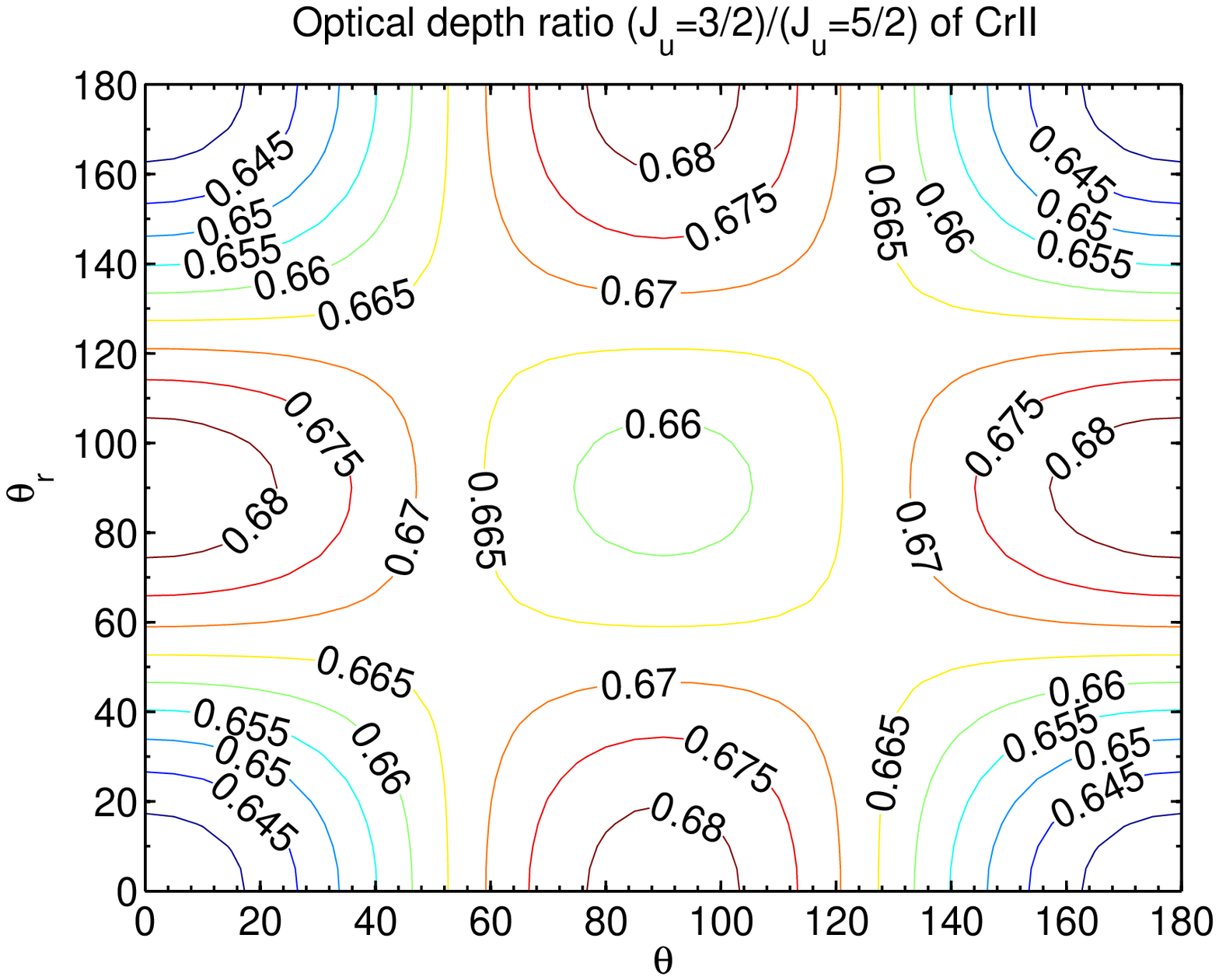}
\includegraphics[%
  width=0.45\textwidth,
  height=0.3\textheight]{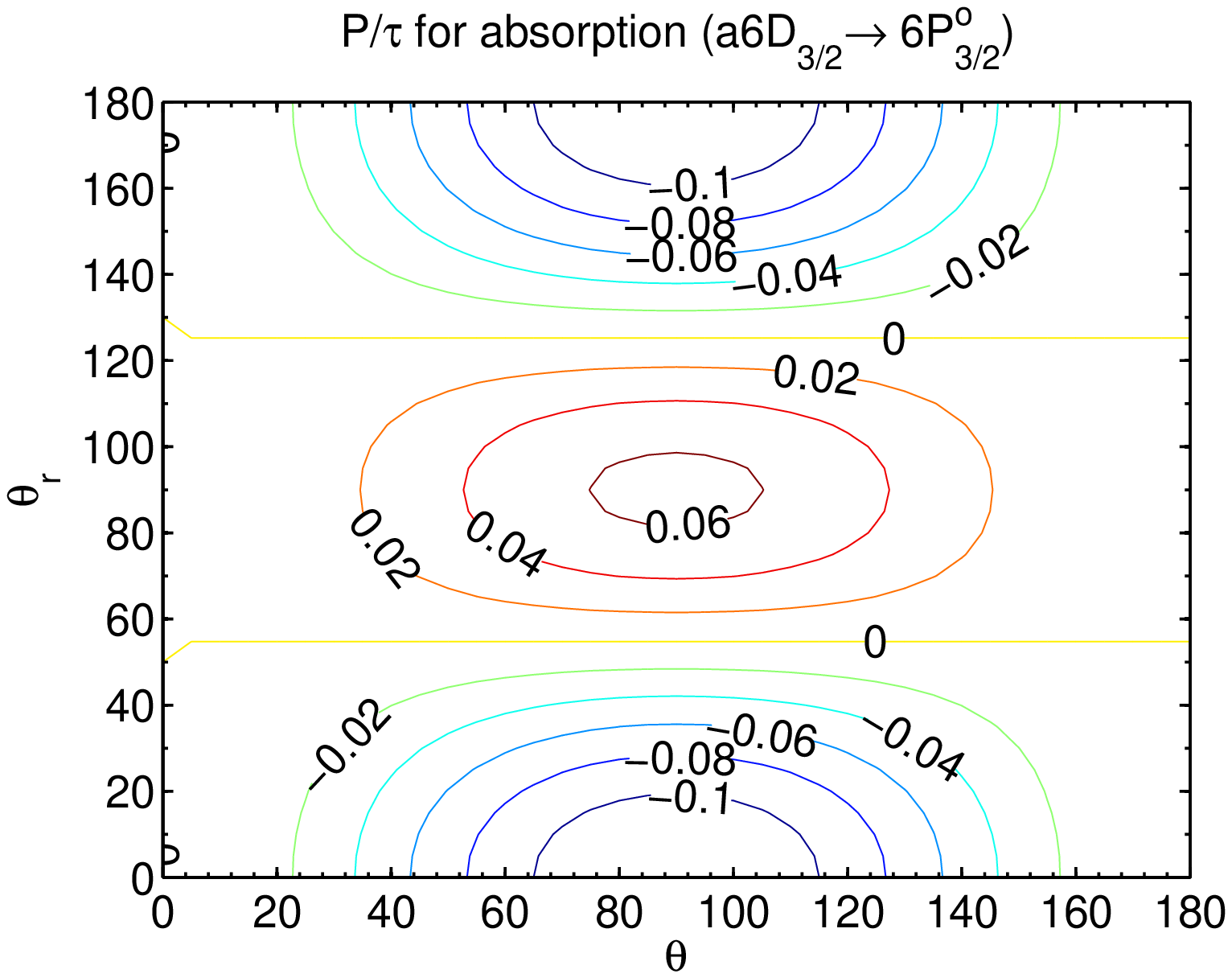}
\includegraphics[%
  width=0.45\textwidth,
  height=0.3\textheight]{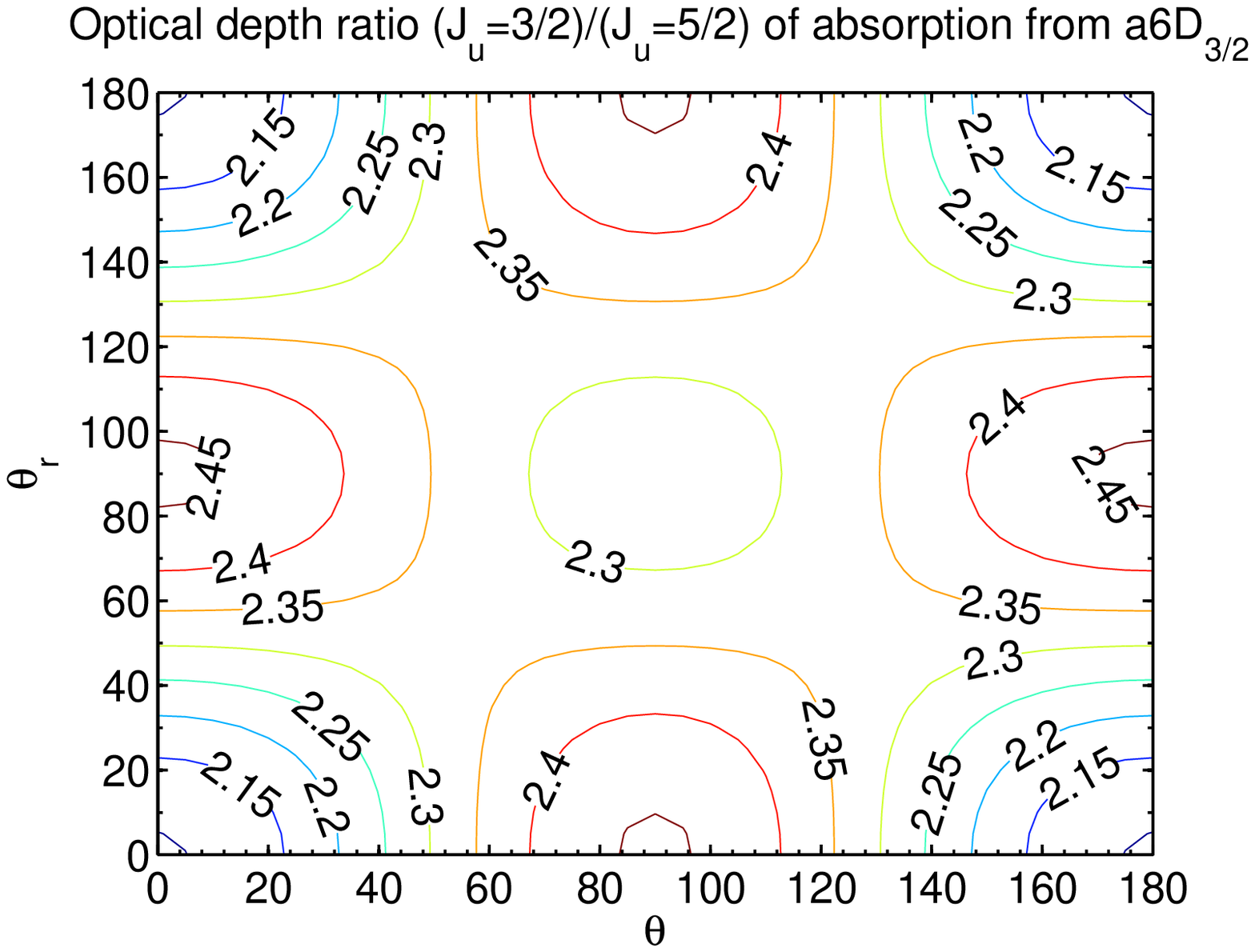}
\caption{The contour graphs of Cr II polarization and line ratio. The upper panels are for the absorption from the ground state $a6S$ and the lower panels are for the absorption from the metastable state $a6D$.}
\label{Cr2contour}
\end{figure}

{\bf Cr II} has a ground state $a6S_{5/2}$ and excited state $z6P^o_{3/2,5/2,7/2}$. Transition between them generates the triplet 2056, 2062, 2066\AA. 
The most important difference between CrII and the other species 
discussed in this paper is the presence of 
a {\it metastable} state $a6D_{1/2,3/2,...,9/2}$ (with lifetime $\tau_T\sim 10$s), which can act as the
proxy of the ground state. The transitions between the metastable
levels and the upper levels are permitted
(with wavelength ranging in 2740-2767\AA). An analogy thus can be made between Cr II and the atomic species with multiple lower levels we deal with in this paper. In the strong pumping regime, both the ground level and the metastable level can be aligned. Using the similar treatment 
(neglecting the  
forbidden transition between the metastable levels and the ground level), we can get the alignment and polarizations for the two groups of transitions (the triplet from the ground
state and the absorptions from the metastable state). Since there are many levels involved and the dimension of the equations we need to solve is $17\times 17$, we only present selected numerical results here. The dipole components of the density matrices for the ground and metastable level is given in Fig.\ref{ground}{\it right}. We show also the polarization of the line $(a6S-z6P^o_{3/2}, 2066\AA) $ and the line $(a6D_{3/2}-z6P^o_{5/2}, 2750\AA$) in Fig.\ref{Cr2contour}.

There is an advantage to use the absorption lines from the metastable level. Because the energy of the metastable level is high, they can be populated only through optical pumping. The absorption lines from the metastable levels therefore only traces the medium in the vicinity of a pumping source. Absorption from the ground state, however, can happen in regions much further away where alignment is marginal. The actual observed signals shall depend on $N_{a}/N_{tot}$, the ratio of column density to total column density.  Combining the absorptions from the metastable state and the ground state, allows polarimetric testing of the conditions in the vicinity of the pumping source. Because of that we intend
to provide a study of more atoms with metastable states (see Table~\ref{ch3t2})
in our next paper.

\subsubsection{Weak Pumping}
\label{weakp}
Now let us consider the second regime in which the radiation source is very distant so that the magnetic dipole transition rate within the ground state dominates over the optical pumping rate. Because of the faster magnetic dipole decay, we can neglect the optical pumping from levels other than the ground level, {\it i.e.}, only the ground level is alignable. In this case, additional magnetic dipole transitional terms should be added to Eq.(\ref{evolutiong}). For the ground level,
\bea
\dot{\rho^k_0}(J^0_l)&=& \sum_{J_l}(-1)^{J^0_l+J_l+k+1}[J_l]\left\{\begin{array}{ccc}
J^0_l & J^0_l & k\\J_l & J_l &1\end{array}\right\}A_m(J_l\rightarrow J^0_l)\rho^k_0(J_l) \nonumber \\
&&\sum_{J_u}(-1)^{J^0_l+J_u+k+1}[J_u]\left\{\begin{array}{ccc}
J^0_l & J^0_l & k\\J_u & J_u &1\end{array}\right\}A(J_u\rightarrow J^0_l)\rho^k_0(J_u) \nonumber \\
&-&[J^0_l]\sum_{J_uKk'}B(J^0_l\rightarrow J_u)(-1)^{J^0_l-J_u+k+k'+1}(3[k,k',K])^{1/2}\nonumber \\
&&\left\{\begin{array}{ccc} 
1 & 1 & K\\J^0_l& J^0_l & J_u\end{array}\right\}\left\{\begin{array}{ccc} 
k & k' & K\\J^0_l& J^0_l & J^0_l\end{array}\right\}\left(\begin{array}{ccc}
k & k' & K\\ 0 & 0& 0\end{array}\right)\bar{J}^K_0 \rho^{k'}_{0}(J^0_l),
\label{wpevolutiong}
\eea
For those lower levels other than the ground level, 
\bea
\dot{\rho^k_0}(J_l)&=& -\sum_{J'_l}A_m(J_l\rightarrow J'_l|_{E(J)>E(J')})\rho^k_0(J_l)+\sum_{J'_l}(-1)^{J'_l+J_l+k+1}[J'_l]\left\{\begin{array}{ccc}
J_l & J_l & k\\J'_l & J'_l &1\end{array}\right\}A_m(J'_l\rightarrow J_l|_{E(J')>E(J)})\rho^k_0(J'_l)\nonumber\\
&+&\sum_{J_u}(-1)^{J_l+J_u+k+1}[J_u]\left\{\begin{array}{ccc}
J_l & J_l & k\\J_u & J_u &1\end{array}\right\}A(J_u\rightarrow J_l)\rho^k_0(J_u)
\eea
As discussed earlier, because of fast magnetic mixing, there are no interference terms and thus there are only $\rho^k_0$ terms in the above equations.
Combining with Eq.(\ref{evolution}), we get the following set of linear equations:
\bea
&&\sum_{J_u}\frac{A(J_u\rightarrow J^0_l)}{\sum_{J'_l} A(J_u\rightarrow J'_l)}[J^0_l]B(J^0_l\rightarrow J_u) \sum_{k'}\rho^{k'}_{0}(J^0_l)\sum_{K}(-1)^{k+k'}\nonumber \\
&&(3[k,k',K])^{1/2}\left(\begin{array}{ccc}
k & k' & K\\ 0 & 0& 0\end{array}\right)\bar{J}^K_0\left[(-1)^{J^0_l+J_u}[J_u]\left\{\begin{array}{ccc}
J^0_l & J^0_l & k\\J_u & J_u &1\end{array}\right\}\left\{\begin{array}{ccc} 
1 & J_u & J^0_l\\1& J_u & J^0_l\\ K &k& k' \end{array}\right\}+\right. \nonumber \\
& &\left.(-1)^{J^0_l-J_u}\left\{\begin{array}{ccc} 
1 & 1 & K\\J^0_l& J^0_l & J_u\end{array}\right\}\left\{\begin{array}{ccc} 
k & k' & K\\J^0_l& J^0_l & J^0_l\end{array}\right\}\right]\nonumber\\
&+&\sum_{J'_l}\delta_{J^0_l\pm1, J'_l}(-1)^{J^0_l+J'_l+k}[J'_l]\left\{\begin{array}{ccc}
J^0_l & J^0_l & k\\J'_l & J'_l &1\end{array}\right\}A_m(J'_l\rightarrow J^0_l)\rho^k_0(J'_l) =0
\label{ismJl0}
\eea
\bea
&&\sum_{J_u}\frac{A(J_u\rightarrow J_l)}{\sum_{J'_l} A(J_u\rightarrow J'_l)}[J^0_l]B(J^0_l\rightarrow J_u) \sum_{k'}\rho^{k'}_{0}(J^0_l)\sum_{K}(-1)^{k+k'}(3[k,k',K])^{1/2}\bar{J}^K_0\nonumber \\
&&\left(\begin{array}{ccc}
k & k' & K\\ 0 & 0& 0\end{array}\right)\left[(-1)^{J^0_l+J_u}[J_u]\left\{\begin{array}{ccc}
J^0_l & J^0_l & k\\J_u & J_u &1\end{array}\right\}\left\{\begin{array}{ccc} 
1 & J_u & J^0_l\\1& J_u & J^0_l\\ K &k& k' \end{array}\right\}+\sum_{J'_l}A_m(J_l\rightarrow J'_l|_{E(J'_l)<E(J_l)})\rho^k_0(J_l) \right. \nonumber \\
&+&(-1)^{J_l+J'_l+k}[J'_l]\left\{\begin{array}{ccc}
J_l & J_l & k\\J'_l & J'_l &1\end{array}\right\}A_m(J'_l|_{E(J'_l)>E(J_l)}\rightarrow J_l)\rho^k_0(J'_l)=0
\label{ismJl}
\eea
By solving these equations, we can get the ground level ($3P_2$) dipole density component of  {\bf O I} and {\bf S I}:
{\scriptsize
\bea
\sigma^2_0(J^0_l=2)&=&
\left[\left(5.14D23^2+2.21
   P22 D23+13.21 S21
   D23-44.74 S21^2+9.96
   P22 S21\right) \cos
   ^4\theta_r+\left(-46.47
   D23^2-63.20 P22
   D23+31.17 S21
   D23\right.\right.\\
&-&\left.\left.18.59 P22^2+113.35
   S21^2+58.30 P22 S21\right) \cos
   ^2\theta_r+14.92
   D23^2+6.20 P22^2-32.81
   S21^2+20.82 D23
   P22-11.86 D23
   S21-20.54 P22
   S21\right]/\nonumber\\
&&\left[\left(-2.35
   D23^2+5.63 P22
   D23-5.66 S21
   D23+5.45 P22^2+11.86
   S21^2+2.77 P22 S21\right) \cos
   ^4\theta_r+\left(21.83
   D23^2+8.01 P22
   D23+47.35 S21
   D23\right.\right.\nonumber\\
&-&\left.\left.22.10 P22^2+37.97
   S21^2+40.71 P22 S21\right) \cos
   ^2\theta_r-71.97
   D23^2-141.15
   P22^2-133.61
   S21^2-215.84D23
   P22-199.53 D23
   S21-281.43 P22 S21\right]\nonumber
\eea}
where S21, P22, D23 are $= \sum I_\nu S_\nu$. For O I,
\bea
S21&=&I_S S_{1302.2}=1.08I_S=I_{1302.2}S_{1302.2}+ I_{1039.2}S_{1039.2}, \, P22=0,\nonumber\\
D23&=&I_D S_{988.7}=0.732I_D=I_{1025}S_{1025}+ I_{988.7}S_{988.7}+I_{971.7}S_{971.7}.
\label{O1}
\eea 
for S I,
\bea
S21&=&I_{S}S_{1826.2}=0.502 I_{1826.2},\, P22=I_{P}S_{1295.6}=1.78 I_{1295.6}, \nonumber\\
D23&=&I_DS_{1425}=0.149 I_D=I_{1474}S_{1474}+I_{1425}S_{1425}+I_{1316.5}S_{1316.5}
\label{S1}
\eea
The line strengths $S_{\nu}$ in the above two equations are taken from NIST Atomic Spectra Database. The line intensities for various environments are listed in Table 5, where the values for interstellar radiation field (ISRF)are particularly relevent to the weak pumping regime we discuss here.

We only provide the density matrices for the ground level. Absorption from the levels other than the ground level is negligible because the absorption rate is less then the magnetic decay rate.

\subsubsection{Role of collisions}
\label{collision}
In the above calculations, collisions are neglected. Now let us see whether this is a reasonable assumption. Basically there are two types of collisions. The first type is an inelastic collision, which excites transitions among the J levels of the ground state. The second type of collision is elastic and only causes the angular momentum {\bf J} to flip. 

The inelastic collision can happen with electrons or hydrogen atoms. Both rates can be found in Spitzer (1978). The transition rate for collisions with electrons is  
\be
f_{kj}=n_e\gamma_{kj}=\frac{8.63\times 10^{-6}n_e\alpha(j,k)}{[k]T^{1/2}}{\rm cm}^3{\rm s}^{-1},
\ee
where $[k]$ is the statistical weight of level $k$. $\alpha$, the collisional strength, is of order unity (Osterbrock 1974), while the magnetic dipole emission rate between the lower levels is $A_m\gtrsim 10^{-6}{\rm s}^{-1}$. Apparently, the collisional rate is much smaller for diffuse medium. For collisions with hydrogen atoms, $f_{kj}=n_H\gamma_{kj}\lesssim 10^{-10}n_H{\rm s}^{-1}$; this is also smaller than the magnetic dipole decay rate $A_m$. Thus both collisions with electrons and hydrogen atoms are less frequent than magnetic dipole decay in diffuse medium.

The second collision involves spin flip and is dominated by Van der Waals interactions with neutral atoms, in particular, hydrogen atoms. The corresponding rate is (Lam \& ter Haar 1971, Landi Degl'Innocenti 2004),

\be
f_{sf}\simeq 2\times 10^{-10}\left(\frac{13.6{\rm eV}}{E_i-E_J}\right)^{0.8}\left[T(1+\frac{1}{\mu})\right]^{0.3}n_H {\rm s}^{-1},
\ee
where $E_i$ is the ionization potential, $E_J$ is the energy of $J$ level, $\mu$ is the atomic weight.

In the weak pumping regime which we discussed in \S\ref{weakp}, collisions can become comparable to the optical pumping rate. In this case alignment is partially destroyed\footnote{To calculate the residual alignment, one should include collisional transitions in the statistical equilibrium equations of the atomic populations.}. Since pumping rates for different species are different, we can find the tomography of the magnetic field by utilizing lines from various species. By equating the collisional rate max$(f_{kj}, f_{sf})$ to the optical pumping rate, we obtain the alignable range for an O type star,
\be
r_2 \simeq \cases{16\left(\frac{0.1{\rm cm}^{-3}}{n_e}\sqrt{\frac{T}{8000{\rm K}}}\right)^{1/2}{\rm pc},& for C II 
\cr
 12.6\left(\frac{0.1{\rm cm}^{-3}}{n_e}\sqrt{\frac{T}{8000{\rm K}}}\right)^{1/2}{\rm pc},& for Si II \cr}.
\label{r0}
\ee
For a typical QSO with luminosity $<L>=\lambda L_\lambda=10^{46.16}{\rm ergs\cdot s}^{-1}$ at 1100\AA and a spectrum (Telfer et al. 2002) 
\bea
F_\nu \propto \left\{\begin{array}{rl}\nu^{-0.7},& (\lambda>1250\AA)\\
\nu^{-1.7},& (\lambda<1250\AA)\end{array}\right.,
\eea
the alignable range would be 
\bea
r_2 \simeq \left\{\begin{array}{rl}27.2\left(\frac{10^{-4}{\rm cm}^{-3}}{n_e}\sqrt{\frac{T}{10^6{\rm K}}}\right)^{1/2}{\rm Mpc},& {\rm for ~\,C\,~ II}\\
11.6\left(\frac{10^{-4}{\rm cm}^{-3}}{n_e}\sqrt{\frac{T}{10^6{\rm K}}}\right)^{1/2}{\rm Mpc},& {\rm for\,~ Si\,~ II}\end{array}\right. 
\eea

In actual observations, there is inevitably an average along line of sight. The absorption from atoms outside the alignable range tend to reduce the degree of polarization. This adds another dependence 
on $N_{a}/N_{tot}$, the ratio of alignable column density to total column density. In this case, there is an advantage to choose species present only near a pumping source. For instance, the column densities
of many ions within an HII region may dominate the column densities
along the line of sight. In this case, the conditions in the  HII region
will be traced.

\section{Summary of the procedure adopted}

In this paper, we provided calculations for the alignment of atoms with fine structure owing to pumping by anisotropic radiation in a magnetized medium. The angular momentum of the radiation is transferred to the atoms and causes differential occupation in terms of angular momentum (magnetic quantum number) of the ground state. This results in atomic alignment. Any background radiation (including the pumping source) coming through these atoms will be absorbed differentially, causing polarization in absorption lines. 

In most astrophysical environments, incident radiation is axially symmetric. This makes the irreducible density matrix the suitable formalism to describe the system (see App.\ref{irredmatrix}). In this case, alignment is simply represented by the $\rho^2_q$ term, which causes linear polarization. Total population is given by the $\rho^0_0$ term. Since the incident radiation is unpolarized, there is no orientation, and the corresponding terms $\rho^1_q$ are zeros accordingly. Circular polarization, resulting from orientation, does not exist in this case\footnote{If the background source is polarized in a
 different direction with alignment,
circular polarization can be 
generated due to dephasing. This is a second order effect, and we shall discuss it in our companion paper.}. 

When the magnetic field exists, it is convenient to choose the magnetic field direction as the quantization axis. Coherence terms $\rho^K_{Q\neq 0}=0$  disappear because of fast magnetic mixing on the ground state. In this sense, atoms are realigned along the magnetic field compared to the case without a magnetic field. This requires a transformation from the general observational frame to the theoretical frame where the magnetic field defines the $z$ axis (Fig.\ref{radiageometry}). This can be done by two Euler rotations (see App.\ref{euler}).  

We solve the statistical equilibrium equations of both ground state and excited state and obtain the density matrix $\rho_Q^K$ of the ground states. Multilevel atoms are considered as well. All of the transitions between the levels in ground state and the levels in upper state should be taken into account to get the correct atomic alignment. If there are multiple lower levels and the decay rate between them is larger than the optical pumping rate, the transitions between them should be also taken into account and their corresponding statistical equilibrium equations should be included to solve the level populations. These transitions can substantially change the alignment. This is the weak pumping case we studied in \S\ref{weakp}.

Then we insert the dipole component of the density matrix of the ground state $\sigma^2_0$ into the equation for absorption coefficients (Eq.\ref{Mueller}), yielding line ratios and polarizations for absorption lines (Eqs\ref{absorb},\ref{tauratio}). Polarization of absorption is either parallel or perpendicular to magnetic field in the pictorial plane because of fast precession around the magnetic field in the ground state.

Fig.\ref{diagram}{\it right} illustrates when and in what form we expect to
see atomic alignment depending on the atoms and their environmental
conditions. It also illustrates the computational steps to
determine the direction of magnetic field.

\section{Discussion}
\subsection{Choice of atoms}
We have chosen a particular set of atoms (see Table~\ref{ch3t1}) as examples of alignment for atoms with fine  structures. Since the only requirement for atomic alignment is having a ground level with total angular momentum $J>1/2$, there are many more atomic species that can be explored. Regardless of the differences in specific atomic structures, the physics is similar. By considering transitions from multiple levels to multiple levels, we have developed a procedure that can be applied to virtually all atoms fine structure. We did not include level coherence as it is negligible except for a highly turbulent medium where $\delta v\gtrsim 100$km/s. For multiplets, the absorptions $J_i\geq J_f$ are usually more polarized than other components which provides more routes and more balance between $\sigma$ and $\pi$ transitions.

Practically, however, there is inevitably an averaging along line of sight, which adds another dependence on the ratio of alignable column density and total column density $N_a/N_{tot}$. For different species, this ratio is different. The ratio should be close to 1 for highly ionized species which only exist near radiation sources. The same is true
 for the
absorption from metastable state, for instance, for 
CrII that we discussed earlier. The energies between
the upper and metastable state in this case are too high to be affected by collisions with thermal particles in cool phases along the line of 
sight. Thus the corresponding absorption lines are also mostly 
affected by the medium in the vicinity of a star (or QSOs). 
Combining different species (with different $N_a/N_{tot}$), it is possible to get a tomography of the magnetic field {\it in situ}. To
extend the technique we shall present elsewhere calculations for
more atoms with metastable states. 

The particular choice of atoms to use depends both on the instruments
available and the object to be studied. With the advent of space-based
UV polarimetry many more alignable transitions will become important.
This will allow detailed studies of magnetic fields in circumstellar
regions, diffuse ISM, intergalactic medium, supernovae 
remnants\footnote{This possibility was mentioned to us by Ken Nordsieck.},QSOs etc. However, even at present a 
substantial progress may be achieved using ground-based and
possibly sounding rocket observations.


\subsection{Wavelengths employed and particular cases}

The incomplete list of astrophysically interesting transitions in 
Table~\ref{ch3t1},\ref{ch3t2} shows that many of the transitions are in the  
UV part of the spectrum. At the moment, the optical polarization is easily available, e.g., Cr I, Ti II. Furthermore, we expect that in the years to come, 
UV polarimetry will become an essential tool for astrophysical
research. In the mean time, a number of transitions (Ti II, Cr I, Fe I, etc.) as well
as atomic transitions shifted due to cosmological redshifts can be studied using present ground based polarimetry
facilities. The latter is applicable to, e.g., Cr II detectable
in QSO absorption (Kulkarni et al. 2005).

We did not discuss transitions beyond the Lyman limit 912\AA.  There are actually many lines in this range, especially for absorptions in QSOs (Verner, Barthel \& Tytler 1994) and Lyman $\alpha$ clouds (see Akerman et al. 2005). Indeed for diffuse ISM, absorptions are severe. However, for QSOs, this constraint is relieved because of proximity effect, which is the measured decrease in the number of Ly$\alpha$ clouds induced by the UV radiation from a QSO in its vicinity (Madau 1992). When the light reaches those Ly$\alpha$ clouds far apart, it is already red-shifted to a wavelength longer than 912$\AA$ so that it is free of the absorptions from H I as well. Similarly these lines are also informative for the local bubble (see Redfield \& Linsky 2004).

 As we discussed for the case of weak pumping (\S\ref{weakp}), the forbidden decay
of the ground state may result in the loss of alignment for atoms like C II and Si II, which is an interesting observable effect in itself. For other transitions (e.g., O I, S I), the decay does not erase but modifies the alignment. 


It is worth mentioning
 that when the photon arrival rate is comparable the
Larmor frequency, the polarization becomes sensitive to the amplitude of
magnetic field. This opens an avenue for obtaining the {\it strength}
of weak magnetic fields, especially for circumstellar region and QSOs. If there are several species spatially 
correlated, it is possible to measure both the magnetic field strength and direction. In addition, even measurements of
 unpolarized lines are useful in
conjunction with studies of aligned species. If the photon arrival rate is less than Larmor precession rate, atoms are aligned with respect to the radiation field and their line polarization does not contain information about the magnetic field. For instance as we pointed out in \S\ref{relscales}, magnetic field can only realign CII at a distance $\gtrsim 7.7$Au from an O star if the magnetic field strength $\sim 5\mu$G.  

We would like to emphasize here that we deal with optically thin case, or unsaturated absorption in this paper. radiative transfer needs to be considered for the saturated regime. 

\subsection{Utility of transitions}

{\it There are a number of parameters that determine the polarization}: the direction of the magnetic field and radiation field ($\theta, \theta_r$), the percentage of radiation anisotropy $W_a/W$ and the ratio of alignable column density and total column density $N_a/N_{tot}$. 
 The list of observables we have are the degree of polarization and the line intensity ratios (equal to optical depth ratios for optically thin case) for multiple lines. Since there are a lot of alignable atomic species,  we can have more known observables to determine the magnetic field if we make sure that these atomic species coexist. One can also use the transitions
between different levels of the same atom.


Combining different atomic species, we can make a tomography of magnetic field as the ratio $N_{a}/N_{tot}$ varies with each atomic species. Moreover, the fact that the radius of alignment changes for different
environments, allows polarimetric testing of the conditions in the vicinity of the pumping source.

If the alignment is produced by the background radiation source
which is used to study the absorption lines\footnote{We call this
case a ``degenerate'' case to distinguish from a case where the
background source and the source of the aligning radiation differ.}
(see Fig.~1, right), the line of sight coincides with the direction of radiation, resulting in $\theta_r=\theta$.   

Our calculations in this paper were aimed at exploring
the feasibility of the measurements of the astrophysical magnetic
fields using atomic alignment. In the paper above we
dealt with the fine structure line transitions, while the hyperfine
structure transitions are considered in a separate paper. Taken together,
the alignable species allow studies of magnetic fields and environmental
conditions for various astrophysical systems, from comet atmospheres
 to stellar winds, supernovae, reflection nebulae, interstellar gas and the intergalactic medium.

\subsection{Studies using contour plots}

Practically, we can make use of contour plots like what we did for S II, Cr II to get information on the magnetic field and environment from observational data. For instance, if one measures for S II ($J_u=1/2$) absorption line degree of polarization per unit optical depth 8\% and the equivalent width ratio of S II ($J_u=1/2,3/2$) doublet 0.48\footnote{For optically thin lines, the equivalent width ratio is equal to optical depth ratio as given in the lower panels of Fig.\ref{S2contour}}. First from the direction of polarization, we know that the magnetic field projection is either parallel or perpendicular to the direction of polarization in pictorial plane (azimuthal angle $\phi_B$, see Fig.\ref{radiageometry}). This is already the whole information one can expect from the Goldreich-Kylafis effect (Goldreich \& Kylafis 1982). Then, using the degree of polarization and the line ratio, one can get further information about the magnetic field in the plane of line of sight $\theta$, and also the direction of pumping source $\theta_r$. By overlapping the two contour graphs (Fig.\ref{S2contour}), we can find their crosses at $\theta_r\simeq 90^o\pm 72^o$ and $\theta\simeq 90^o\pm 7^o$\footnote{The uncertainties come from the degeneracies of $\sin^2\theta$ and $\cos^2\theta_r$ in Eqs(\ref{absorb},\ref{tauratio}).}. With $\theta_r$ known, we can easily remove the $90^o$ degree degeneracy of $\phi_B$ according to the Van-Vleck effect and thus obtain the 3D information of the magnetic field. If we know two species coexist spatially, we can also use their ratio of degrees of polarizations and their ratio of equivalent width. In this case, we don't need to know the optical depth for individual line. Alternatively, $\theta_r$ and $\theta$ can be obtained by solving Eqs(\ref{absorb},\ref{tauratio}). Calculating $\theta_r$ and $\theta$ several ways cross-check the results and decreases the uncertainties.

\subsection{Complementary techniques}

Atomic alignment can render unique information about the magnetic field.
For instance, grain alignment\footnote{Grains have tendency to be aligned 
with their long
axes perpendicular to the magnetic field. Curiously enough, the interaction with photons
is also important for grain alignment.} (see review by Lazarian
2003) provides information about the component in the plane of sky. 
Similar to the case of interstellar dust, the rapid precession of atoms
in a magnetic field makes the direction of polarization sensitive to the
direction of the underlying magnetic field. As the precession of magnetic moments
of atoms is much faster than the precession of magnetic moments of grains, 
atoms can reflect much more rapid variations of magnetic field.  The limited
radius of atomic alignment is also an advantage to enable local measurement of
3D magnetic fields while grain alignment tests global 2D topology. More 
importantly, alignable atoms and ions can reflect magnetic fields in the
environments where either the properties of dust change or the dust cannot survive.
This opens wide avenues for magnetic field research in circumstellar regions,
interstellar medium, interplanetary medium, intracluster medium and quasars, etc.

The complementary nature of atomic alignment
to the Goldreich-Kylafis effect\footnote{Atomic alignment has some similarity to
 Goldreich-Kylafis effect, which also measures magnetic field through 
magnetic mixing. 
However, Goldreich-Kylafis effect deals with the polarization of radio 
lines. 
The upper states of radio lines are so long lived that significant magnetic 
mixing can happen 
among  different magnetic sublevels of these states. 
Atomic alignment, on the other hand, happens with ground states 
of optical and UV transitions.} (Goldreich \& Kylafis 1982; Girart, Crutcher, \& Rao 1999). is that the former is applicable to the
regions near emitting sources (e.g. circumstellar regions), while the
latter is applicable to molecular clouds.
 Moreover, atomic alignment provides 3D tomography of magnetic fields that is not measurable otherwise.

A detailed discussion of alignment of atoms in different conditions 
corresponding to 
circumstellar regions (see Dinerstein, Sterling \& Bowers 2006), AGNs (see Kriss 2006), Lyman alpha clouds (see Akerman et al. 2005), interplanetary space (see Cremonese et al. 2002), the local bubble (see Redfield \& Linsky 2002, 2004), etc., is given in our companion paper. 
Note, that for interplanetary studies,
one can investigate not only spatial, but also temporal variations
of magnetic fields. This can allow cost effective way of studying
interplanetary magnetic turbulence at different scales.

The change of the optical depth is another important consequence of atomic alignment. The actual calculations should take into account the realignment caused by a magnetic field. It may happen that the variations of the optical depth caused by
alignment can be related to the Tiny-Scale Atomic Structures (TSAS) observed
in different phases of interstellar gas (see Heiles 1997).

\section{Summary}

In this paper we identified a set of atoms and
ions with fine structure that
provide substantially intensive absorption lines in
the visible and near UV wavelength range.  
We calculated the degree of alignment of these species and obtained the
relation between the Stokes parameters of the absorbed
radiation and the 3D orientation of magnetic field with respect to the observer and the pumping/absorbed sources.
Our calculations can easily be extended to other absorbing  species
with fine structure. In particular, we have shown that:

1. The ground-state alignment happens as the result of 
interaction of atoms and ions 
that have $>2$ sublevels on the ground state
with the anisotropic flow of photons. If the rate of excitation from
the ground state is lower than the rate of atomic precession in the
external magnetic field, the alignment happens with respect to the magnetic
field. 

2. The ground-state alignment affects the polarization state of the
 absorbed light as well as the intensity of different
line components. Combining information from different components
and absorption lines one can study both 3D topology
of magnetic fields and physical conditions
in important astrophysical environments that include
quasars, interstellar medium and extragalactic gas, etc.

3. For the light absorbed by aligned atoms,  the direction of 
polarization
is either parallel or perpendicular to the magnetic field in the plane of sky. The switch between
the two possibilities happens at the Van-Vleck angle between the direction of the magnetic field and the pumping radiation.

\begin{acknowledgments}
We are grateful to Ken Nordsieck for fruitful discussions and providing
 relevant data. We thank J. Everett for reading our manuscript and giving helpful comments. We thank the referee for valuable comments and suggestions. Our thanks also go to N. Dalal and P. McDonald for their help and input. HY acknowledges the current support by CITA and the National Science and Engineering Research Council of Canada as
well as the earlier support by NSF grant AST 0098597. AL acknowledges
the support by the NSF grant AST 0098597 and the NSF Center for Magnetic Self-Organization in Laboratory and Astrophysical Plasmas. 

\end{acknowledgments}

\appendix

\section{Irreducible density matrix}
\label{irredmatrix}
In quantum mechanics, a particle (atom or photon) can be described by its wave function $|\psi>=\sum_{JM}c_{JM}|JM>$. An equivalent but more elegant description without the phase arbitrariness is a density matrix $\rho=|\psi><\psi|=\sum_{JM,J'M'}c_{JM}c^*_{J'M'}|JM><J'M'|$. The diagonal term $JM|\rho|JM>=|c_{JM}|^2$ represents the probability of finding the particle on state $|JM>$; off-diagonal terms $<JM|\rho|J'M'>=c_{JM}c^*_{J'M'}$ characterize the coherence between two different states $|JM>$ and $|J'M'>$. The advantage of the description by density matrix becomes apparent when we are dealing with an ensemble of n particles without mutual interactions. In this incident, the density matrix of the ensemble is simply the arithmetic mean of individual densities $\frac{1}{n}\sum_{i=1}^n\rho(i)$.

When magnetic field is present, it is convenient to choose the magnetic field direction as the quantization axis for an atom because $M_J$ remains a good quantum number on the ground state. Meanwhile, the density matrix of radiation (or their classical equivalent Stokes parameters,  see App.\ref{Stokes}) should be obtained in a basis where the line of sight is the quantization axis. This requires several rotations of the successive reference frames (see App.\ref{euler}). Usually, the incident radiation is axially symmetric, while the spontaneous emission is spherically symmetric. This makes the irreducible tensorial operators $\rho^K_Q$ the suitable formalism for performing the calculations (Fano 1957, Dyakanov \& Perel 1965, Omont 1965, Bommier \& Sahal-Brechot 1978). The relation between irreducible tensor and the standard density matrix is 
\be
\rho^K_Q(J,J')=\sum_{MM'}(-1)^{J-M}(2K+1)^{1/2}\left(\begin{array}{ccc}
 J & K & J'\\
-M & Q & M'\end{array}\right)<JM|\rho|J'M'>.
\label{irreducerho}
\ee
For photons, we shall consider only their electric dipole part, which has a spin $J\equiv 1$. The irreducible tensor has a clear physical meaning. In particular, $\rho_0^0$ is proportional to the overall population of an atomic level or the total intensity of the radiation;  $\rho^1_0$ describes orientation of an atom or circular polarization of radiation; $\rho^2_Q$ represents the alignment term, which are responsible for linear polarization; the terms with $Q\neq 0$ are coherence terms. For atoms, $\rho^K_Q(J,J')$ is the level-crossing coherence (Bommier \& Sahal-Brechot 1978), $\rho^K_{Q\neq 0}(J)$  is the Zeeman coherence as they correspond to the crossing term $\rho_{MM'}$ according to the selection rule of 3j symbol (App.\ref{angmtmcpl}) in Eq.(\ref{irreducerho}). Note that in the regime we study (namely, weak magnetic field [$\lesssim 1$G] and low temperature diffuse medium), individual J levels are much narrower than their separations so that level-crossing coherence is negligible.

\section{Radiative transitions}
\label{radiatheory}
We consider atoms with {\it fine} structure. The ground level of the atoms has an angular momentum $J,M$ and the excited state is $J',M'$\footnote{Throughout this paper, we follow the convention that $M$ represents the projection of angular momentum along magnetic field--the quantization axis.}. For spontaneous emission from a state $J'M'$ of higher energy to a lower energy state $J,M$, the transition probability per unit time is
\be
a=\frac{64\pi^4e^2a_0^2\nu^3}{3hc^3}\sum_q |<JM| V_{q} |J'M'>|^2
\label{smalla}
\ee
where $V_q^{i,o}={\bf r\cdot e}_q^{i,o}$ is the projection of the dipole moment 
along the basis vector ${\bf e}_q$ of radiation, ${\bf e_{\pm}}= (\mp{\bf \hat x}-{\bf \hat y})/\sqrt{2}$, 
${\bf e}_0={\bf \hat z}$. According to the Wigner-Eckart theorem (Weissbluth 1978), the electrical dipole matrix element is
\begin{eqnarray}
R_{J'J}=<JM| V_{q} |J'M'>=(-1)^{J'+M-1}\left(\begin{array}{ccc}J &1& J'\\-M &q& M'\end{array}\right)<J||V_q||J'>,
\label{fine}
\end{eqnarray}
where the matrix with big "( )" is Wigner 3j symbol, which is a measurement of coupling of ground angular momentum ($J,M$) and photon angular momentum ($1,q$) forming angular momentum ($J',M'$) of the upper level (see App. A). $<J||V_q||J'>$ is the square root of the line strength, whose value can be found in atomic physics books, e.g., Cowan (1981).

\section{Coupling of angular momentum vectors}
\label{angmtmcpl}
Each atomic (or photon) state has a particular angular momentum associated with it. Because of its vector nature, we can describe it with two quantum numbers $|jm>$, where $j$ is total angular momentum and $m$ is its projection along the quantization axis. Whenever discussing any atomic processes involving direction, we would find it is inevitable to involve  angles and momenta. Angular momentum theory arises as a result.

When two angular momenta $J, m_1$ and $J', m_2$ are coupled, the resultant angular momentum $j, m$ can be described as (Zare 1988)
 
\be
|jm>=\sum_{m_1, m_2}<Jm_1, J'm_2|jm>|Jm_1, J'm_2>,
\ee
where the elements of transformation $<Jm_1, J'm_2|jm>$ are called Clebsch-Gorden coefficients or Wigner coefficients. To better reveal its symmetric properties, the
Wigner 3-j symbol was introduced
\be
\left(\begin{array}{ccc}J& J' & j_3\\m_1 & m_2 & -m_3\end{array}\right)\equiv (-1)^{J-J'-m_3}[j_3]^{-1/2}<Jm_1, J'm_2|jm>
\ee
The three $J$ must satisfy the triangle relations:
\be
J+J'\geq j_3,\, J'+j_3\geq J, \, J+j_3\geq J'.
\ee
and $m_1+m_2-m_3=0$. If one is concerned radiative transitions, one of the angular momentum ($J',m_2$) would be that of photon (1,q). It then follows that the selection rules for transitions between states $J,m_1$, $j_3,m_3$ are
\be
\Delta j=j_3-J=0,\pm 1, \Delta m=m_3-m_1=0,\pm 1
\ee
with the restriction that $J=j_3=0$ is not allowed.

When three angular momenta $J, m_1\,J',m_2\,j_3, m_3$ need to be coupled, there is more than one way to do this. We can first couple $J, m_1$ and $J', m_2$ to get $j_{12}$, then add $j_3$; or we can first couple $j_3, m_3$ and $J', m_2$ and get the same result. The transformation between $<j_{12}j_3 jm>$ and $<Jj_{23}>$ is given by (Zare 1988)
\be 
|Jj_{23}jm>=\sum_{j_{12}}<j_{12}j_3 j|Jj_{23}j>|j_{12}j_3 jm>
\ee
where the expansion coefficients $<j_{12}j_3 j|Jj_{23}j>$  are called recoupling coefficients as they describe the transformation between two coupling schemes $<j_{12}j_3 jm>$ and $<Jj_{23}>$. A closely related quantity was then defined by Wigner (1965), which has higher symmetry:
\be
\left\{\begin{array}{ccc}J& J' & j_{12}\\j_3& j&j_{23}\end{array}\right\}=(-1)^{J+J'+j_3+j}[j_{12},j_{23}]^{-1/2}<j_{12}j_3 j|Jj_{23}j>
\ee
 
Similarly, the recoupling of four angular momenta leads to the 9-j symbol:

\be
\left\{\begin{array}{ccc}J& J' & j_{12}\\j_4&j_3&j_{34}\\j_{14}&j_{23}&j\end{array}\right\}=[j_{12},j_{23},j_{34},j_{14}]^{-1/2}<j_{12}j_{34} j|j_{14}j_{23}j>
\ee

\section{Stokes Parameters}
\label{Stokes}
The electric field in the plane perpendicular to the propagation direction can be developed over the usual basis $({\bf e}_x,\, {\bf e}_y)$:
\be
{\bf E}(t)=[E_1 \hat{e}_x (\cos\omega t-\delta_1)+E_2 \hat{e}_y (\cos\omega t-\delta_2)]
\ee
When $\delta_1=\delta_2$, the light is linearly polarized. 
For theoretical calculations, it is more convenient to use basis composed of left- (${\bf e}_-$) and right- (${\bf e}_+$) handed circular vibrations. They are defined as
\be
{\bf e}_{\pm}=[\pm{\bf e}_1+i {\bf e}_2]/\sqrt{2}.
\ee
By using (${\bf e}_+,\,{\bf e}_-$), one has a radiation tensor
\be
{\bf \Pi}=\left(\begin{array}{cc}
< E_+E_+> & < E_+E_->\\<E_-E_+> & <E_-E_->\end{array}\right)=\left(\begin{array}{cc} \Pi_{++} & \Pi_{+-}\\ \Pi_{-+} & \Pi_{--}\end{array}\right),
\ee
where $<>$ denotes an average over time.

The definition of Stokes parameters $S_i=[I,\,Q,\,U,\,V]$ (i=0, 1, 2, 3) and their relation with $\Pi$ is as follows:
\bea
I&=& E_1^2+E_2^2= \Pi_{++}+\Pi_{--},\nonumber\\
Q&=& E_1^2- E_2^2= -(\Pi_{+-}+\Pi_{-+}),\nonumber\\
U&=&2 E_1 E_2\cos(\delta_1-\delta_2)= -i(\Pi_{+-}-\Pi_{-+}),\nonumber\\
V&=&2 E_1 E_2\sin(\delta_1-\delta_2)= \Pi_{++}-\Pi_{--},
\eea
The observable quantities: intensity, degree of polarization p, direction of linear polarization $\chi$ (Fig.\ref{nzplane}{\it right}) have simple relations with Stokes parameters:
\be
I=Tr (\Pi),\,\,\,\,p=\sqrt{Q^2+U^2}/I,\,\,\,\, \sin(2\chi)=U/I,\,\,\,\, \cos(2\chi)=Q/I.
\ee

\section{From observational frame to magnetic frame}
\label{euler}
In real observations, the line of sight is fixed, and the direction of the magnetic field is unknown. Thus a transformation is needed from the observational frame to the theoretical frame where the magnetic field is the quantization axis. This can be done by two Euler rotations as illustrated in Fig.\ref{radiageometry}.  In the original observational coordinate (xyz) system, the direction of radiation is defined as the z axis, and the direction of magnetic field is characterized by polar angles $\theta_r$ and $\phi_B$. First we rotate the whole system by an angle $\phi_B$ about the z-axis, so as to form the second coordinate system $x'y'z'$. The second rotation is from the $z(z')$ axis to the $z''$ axis by an angle  $\theta_r$ about the $y'$-axis.  Mathematically, the two rotations can be fulfilled by multiplying rotation matrices,
\bea
\left[\begin{array}{ccc}
\cos\theta_r&0& -\sin\theta_r \\ 0 &1&0\\ \sin\theta_r &0& \cos\theta_r \end{array}\right]
\left[\begin{array}{ccc}
\cos\phi_B & \sin\phi_B & 0 \\-\sin\phi_B&\cos\phi_B & 0\\0 & 0&1 \end{array}\right]=
\left[\begin{array}{ccc}
\cos\theta_r \cos\phi_B& \cos\theta_r \sin\phi_B & -\sin\theta_r\\
	-\sin\phi_B &		\cos\phi_B			& 0\\
	\sin\theta_r \cos\phi_B& \sin\theta_r \sin\phi_B&	\cos\theta_r \end{array}\right].
\eea

\section{More results of CI, Si I, S III for circumstellar case}\label{CSIS}
For a T=15000K black body pumping source, we have obtained the following results,
 
C I:
{\scriptsize
\be
\begin{array}{l}
\rho^{0}_0(J_l=1)\\
\rho^{2}_0(J_l=1)\\
\rho^{0}_0(J_l=2)\\
\rho^{2}_0(J_l=2)\end{array}=
\left(
\begin{array}{l}
 3.1\cos ^{10}\theta_r-50.8
   \cos ^8\theta_r+327.1 \cos ^6\theta_r-1043.7
   \cos ^4\theta_r-1839.1 \cos
   ^2\theta_r+7806.8
   \\
 0.01\cos ^{12}\theta_r-0.44 \cos
   ^{10}\theta_r-2.39 \cos ^8\theta_r-13.13
   \cos ^6\theta_r-29.63 \cos ^4\theta_r-1367.38
   \cos ^2\theta_r+459.60
   \\
 \cos ^{10}\theta_r-14
   \cos ^8\theta_r+65 \cos ^6\theta_r-30
   \cos ^4\theta_r-3140\cos
   ^2\theta_r+10195
   \\-0.4 \cos
   ^{10}\theta_r-15.7 \cos ^8\theta_r+201.4
   \cos ^6\theta_r-1430.7 \cos ^4\theta_r+4352.5
   \cos ^2\theta_r-1299.1\end{array}
\right)\varrho^0_0
,
\ee}
where $\varrho^0_0=\rho^0_0(J_l=0)/(-0.7 \cos ^{10}\theta_r+11.5
   \cos ^8\theta_r-86.0 \cos ^6\theta_r+548.9
   \cos ^4\theta_r-1746.6 \cos ^2\theta_r+4625.5)$

Si I:
{\scriptsize
\bea
\begin{array}{l}
\rho^{0}_0(J_l=1)\\
\rho^{2}_0(J_l=1)\\
\rho^{0}_0(J_l=2)\\
\rho^{2}_0(J_l=2)\end{array}&=&
\left(
\begin{array}{l}
-0.4 \cos ^{10}\theta_r+7.9
   \cos ^8\theta_r-62.9 \cos ^6\theta_r+211.3
   \cos ^4\theta_r+363.1 \cos
   ^2\theta_r-1516.6
   \\
 -0.013 \cos ^{12}\theta_r+0.451 \cos
   ^{10}\theta_r-2.671\cos ^8\theta_r+18.587
   \cos ^6\theta_r+8.482 \cos ^4\theta_r+150.334
   \cos ^2\theta_r-51.711
   \\-0.1 \cos
   ^{10}\theta_r+1.9 \cos ^8\theta_r-4.9
   \cos ^6\theta_r-18.1 \cos ^4\theta_r+635.4
   \cos ^2\theta_r-1975.1
   \\ -0.02 \cos ^{12}\theta_r+0.36\cos
   ^{10}\theta_r+0.38 \cos ^8\theta_r-23.62
   \cos ^6\theta_r+281.19\cos ^4\theta_r-948.70
   \cos ^2\theta_r+285.86\end{array}
\right)\varrho^0_0,
\eea}
where $\varrho^0_0=\rho^0_0(J_l=0)/(-0.05 \cos
   ^{12}\theta_r+0.38 \cos ^{10}\theta_r-3.23
   \cos ^8\theta_r+10.97\cos ^6\theta_r-79.56
   \cos ^4\theta_r+89.78 \cos ^2\theta_r-895.68)$.

S III:
{\scriptsize
\bea
\begin{array}{l}
\rho^{0}_0(J_l=1)\\
\rho^{2}_0(J_l=1)\\
\rho^{0}_0(J_l=2)\\
\rho^{2}_0(J_l=2)\end{array}&=&
\left(
\begin{array}{l}
 52\cos
   ^{10}\theta_r+30\cos
   ^8\theta_r-5710\cos
   ^6\theta_r+23040\cos
   ^4\theta_r+29549 \cos
   ^2\theta_r-130181\\
   \\
 20\cos
   ^{10}\theta_r-544\cos ^8\theta_r+3562 \cos
   ^6\theta_r+1663\cos ^4\theta_r-393 \cos
   ^2\theta_r-179 \\
-25\cos
   ^{10}\theta_r+187 \cos ^8\theta_r-590 \cos
   ^6\theta_r-1259 \cos ^4\theta_r+54726\cos
   ^2\theta_r-165422
   \\
 19\cos ^{10}\theta_r-416\cos ^8\theta_r+1220
   \cos ^6\theta_r+21954\cos ^4\theta_r-91329
   \cos ^2\theta_r+27964
\end{array}
\right)\varrho^0_0,
\eea}
where $\varrho^0_0=\rho^0_0(J_l=0)/(-3 \cos
   ^{12}\theta_r+93\cos
   ^{10}\theta_r-625 \cos
   ^8\theta_r-425 \cos
   ^6\theta_r-155 \cos
   ^4\theta_r+25678\cos
   ^2\theta_r-78043)$.

\end{document}